\documentclass[12pt]{article}
\usepackage[centertags]{amsmath}
\usepackage{amssymb}
\usepackage{amsthm}
\usepackage{enumerate}

\usepackage[english]{babel}
\usepackage[T1]{fontenc}

\makeatletter
  \def\tagform@#1{\maketag@@@{(#1)\@@italiccorr}}
\makeatother

\begin{document}

\title{Geometry and General Relativity in~the~Groupoid Model with~a~Finite~Structure~Group}

\author{Michael Heller \\
Copernicus Center for Interdisciplinary Studies, Cracow, Poland
\and Tomasz Miller\thanks{Corresponding author. E-mail: T.Miller@mini.pw.edu.pl} \quad \ Leszek Pysiak \quad \ Wies{\l}aw Sasin\\
Faculty of Mathematics and Information Science, \\ Warsaw
University of
Technology \\
ul. Koszykowa 75, 00-662 Warsaw, Poland \\
and Copernicus Center for Interdisciplinary Studies,\\ Cracow, Poland}
\date{\today}
\maketitle

\begin{abstract}

In a~series of papers \cite{HSL97,HS99,HPS05} we proposed a~model unifying general relativity and quantum mechanics. The~idea was to deduce both general relativity and quantum mechanics from a~noncommutative algebra ${\cal A}_{\Gamma}$ defined on a~transformation groupoid $\Gamma $ determined by the~action of the~Lorentz group on the~frame bundle $(E, \pi_M, M)$ over space--time $M$. In the~present work, we construct a~simplified version of the~gravitational sector of this model in which the~Lorentz group is replaced by a~finite group $G$ and the~frame bundle is trivial $E=M\times G$. The~model is fully computable. We define the~Einstein--Hilbert action, with the~help of~which we derive the~generalized vacuum Einstein equations. When the~equations are projected to space--time (giving the~``general relativistic limit''), the~extra terms that appear due to our generalization can be interpreted as ``matter terms'', as in Kaluza--Klein-type models. To illustrate this effect we further simplify the~metric matrix to a~block diagonal form, compute for it the generalized Einstein equations and find two of their ``Friedmann-like'' solutions for the~special case when $G =\mathbb{Z}_2$. One of them gives the~flat Minkowski space--time (which, however, is not static), another, a~hyperbolic, linearly expanding universe.
\end{abstract}

PACS Nos.: 02.40.Gh, 04.50.-h, 04.20.Jb 

\maketitle

\section{Introduction}
In a~series of papers (\cite{HSL97,HS99,HPS05} some others will be cited below) we have proposed a~model unifying general relativity and quantum mechanics.  The~core idea of the~model consists of an~attempt to deduce both general relativity and quantum mechanics from a~noncommutative algebra ${\cal A}_{\Gamma}$ defined on a~transformation groupoid $\Gamma $ determined by the~action of the~Lorentz group  (or one of its subgroups) on the~frame bundle $(E, \pi_M, M)$ over space--time $M$. The gravitational sector of the~model is given by the~noncommutative geometry based on the~algebra ${\cal A}_{\Gamma }$ and a~submodule $V$ of the~module of its derivations ${\rm Der} {\cal A}_{\Gamma}$ (we call the~pair $({\cal A}_{\Gamma }, V\subseteq{\rm Der} {\cal A}_{\Gamma })$ differential algebra). The quantum sector of the~model is given by the~regular representation $\pi :{\cal A}_{\Gamma } \rightarrow \mathcal{B}(\mathcal{H})$ of the~algebra ${\cal A}_{\Gamma }$ on a~bundle $\mathcal{H}$ of Hilbert spaces. It turns out that the~model has a~rich mathematical structure \cite{Heller2}, surprising conceptually unifying power \cite{Heller3}, and throws some light on fundamental problems of physics \cite{Fundamental}.

It would certainly be desirable to look at our model in its complete generality by considering the~``full'' differential algebra $({\cal A}_{\Gamma }, \textnormal{Der} \, {\cal A}_{\Gamma })$. However, this should be preceded by a~systematic study of the~very rich module of all derivations $\textnormal{Der} \, {\cal A}_{\Gamma}$. While this study is in progress, the~aim of the~present work is much more modest. In the~present paper, we investigate  a~``toy version'' of our model, in which the~principal bundle is trivial $E = M \times G$ with the~base space $M$ compact and the~structure group $G$ finite. The~transformation groupoid algebra is thus $C^\infty(M \times G \times G)$, which in turn is isomorphic to the~algebra $\mathbb{M}_n(C^\infty(M)) = C^\infty(M) \otimes \mathbb{M}_n(\mathbb{C})$, where $n$ is the~order of the~group $G$. In this paper, we limit ourselves to the~study of the~gravitational sector of this model. Although extremely simple when compared with the~one studied, for example, in \cite{Heller2, Heller3}, this toy model shows many interesting features; only some of them have been visible in our previous works \cite{finite1, finite2}.

Let us add one more motivation of a~general nature for studying geometrical properties of the~algebras $C^\infty(M) \otimes \mathbb{M}_n(\mathbb{C})$. The~celebrated Gel'fand--Naimark theorem states that every \emph{commutative and unital} $C^\ast$-algebra ${\cal C}$ is isometrically isomorphic to the~algebra $C(X)$ of complex continuous functions on a~compact Hausdorff space $X$. This topological space $X$ can be retrieved from ${\cal C}$ by defining it either as the~set of all maximal ideals of ${\cal C}$ (with the~so-called hull--kernel topology) or as the~set of characters of ${\cal C}$ (with the~so-called Gel'fand topology)\footnote{These two definitions (or rather ways of constructing $X$) are equivalent by Gel'fand--Mazur theorem.}. As is well known, the~very idea of noncommutative geometry relies on this fundamental result \cite{Connes, Madore, Bondia, Dubois-Violette1, Dubois-Violette2, Djemai} . The~notion of a~``noncommutative space'' comes indirectly into being through the~study of noncommutative $C^\ast$-algebras that are now interpreted as algebras of ``functions'' living on this ``virtual space''. In general, this space can no longer be thought of as consisting of points, since a~given noncommutative $C^\ast$-algebra might possess too few maximal ideals as well as too few (or no) characters.

One can thus say that a~$C^\ast$-algebra is an~adequate algebraic object to encode the~notion of a~(possibly ``noncommutative'') set with a~certain \emph{topological} structure. However, in order to study differential geometry and, specifically, general relativity, one needs more, namely a~\emph{smooth} structure and a~\emph{pseudo-Riemannian} structure. The~most important example of an~algebraical object that is able to encode these structures\footnote{At least in the~Riemannian case.} is a~spectral triple, conceived first by Connes \cite{Connes}. However, the~approach we follow does not fall under this scheme, we thus have to look for another method of implementing the~smoothness postulate.

First of all, it is reasonable to assume that ${\cal A}$ is a~dense subalgebra of some $C^\ast$-algebra $({\cal C}, \|\centerdot\|_{\cal C})$. In other words,
$\overline{{\cal A}}^{\|\centerdot\|_{\cal C}} = {\cal C}.$
In this way, in a~commutative case, one can readily employ the~Gel'fand--Naimark result and, after interpreting ${\cal C}$ as a~space $C(X)$ of continuous functions on a~compact topological space $X$, one can also treat ${\cal A}$ as a~dense subalgebra of $C(X)$. The~idea would be to consider functions from $\mathcal{A}$ as smooth by definition. Of course, not every subalgebra of $C(X)$ is able to encode the~smooth structure properly (the subalgebra has to be ``rich enough''). One could, for example, employ techniques of the~theory of differential spaces and demand that this subalgebra be closed with respect to superposition with smooth Euclidean functions \cite{Heller1,Structured}. Unfortunately, it is not clear how to extend this condition to the~noncommutative setting. We thus propose the~following approach.

For a~compact Hausdorff space $X$, the~vector space $C^\infty(X)$ can be equipped with the~locally convex topology induced by the~family of seminorms $\{ \| \centerdot \|_D \}$ defined as
\begin{align*}
\forall \, f \in C^\infty\left(X\right) \quad \| f \|_{D} := \sup\limits_{x \in X} \left| D f (x) \right|,
\end{align*}
\noindent
where $D$ denotes a~\emph{differential operator}, i.e. an~endomorphism of $C^\infty\left(X\right)$ that can be defined as a~linear combination of superpositions of finitely many elements of $\textnormal{Der} \, C^\infty\left(X\right)$. Note that the~identity map $\textnormal{id}_{C^\infty(X)}$, regarded as a~superposition of zero derivations, is also a~differential operator. The~set of all differential operators on $C^\infty\left(X\right)$ can also be obtained as the~universal enveloping algebra of the~Lie algebra $\textnormal{Der} \, C^\infty\left(X\right)$ and as such, will be denoted $U(\textnormal{Der} \, C^\infty\left(X\right))$.

It is well known that the~space $C^\infty\left(X\right)$ with the~topology defined in this way is a~complete topological vector space. It is thus natural to formulate the~following definition.

We say that a~unital differential algebra $({\cal A}, V)$ is a~\emph{smooth pre-}$C^\ast$\emph{-algebra} iff ${\cal A}$ is a~dense subalgebra of a~$C^\ast$-algebra $({\cal C}, \|\centerdot\|)$ complete in the~topology induced by the~family of seminorms $\{ \| \, \centerdot \, \|_{D} \; | \; D \in U(V) \}$ defined through
\begin{align*}
\forall \, D \in U(V) \quad \forall \, a \in {\cal A} \ \quad \| a \|_{D} := \| D a \|.
\end{align*}
Note that this family of seminorms includes the~``base'' norm, because
$$\| \, \centerdot \, \| = \| \, \centerdot \, \|_{\textnormal{id}_{\cal A}}.$$

Let us further observe that the~$C^\ast$-algebra ${\cal C}$ can itself be regarded as a~smooth pre-$C^\ast$-algebra $({\cal C},V:=\{0\})$. However, this also means that the~smooth structure of ${\cal C}$ is trivial (the subject of smooth pre-$C^\ast$-algebras will be addressed in a~forthcoming paper).

Let us finally notice that the~algebra ${\cal A}_n := C^\infty(M) \otimes \mathbb{M}_n(\mathbb{C})$ with the~full module of its derivations is a~smooth pre-$C^*$-algebra. Therefore, by investigating its geometrical properties we in fact study a~basic realization of a~smooth pre-$C^\ast$-algebra-based noncommutative geometry.

The algebras ${\cal A}_n$ have yet another interesting property. Their maximal ideals are of the~form $J_m \otimes \mathbb{M}_n(\mathbb{C})$, where $m \in M$, and $J_m$ is an~ideal of $C^\infty(M)$ consisting of all smooth functions that vanish at $m$. Therefore, there is a~bijection between $M$ and the~set of maximal ideals of the~noncommutative algebra ${\cal A}_n$. One can thus say, that although the~noncommutative space associated with the~algebra ${\cal A}_n$ does not consist of points, it nevertheless allows one to retrieve the~point space $M$ associated with the~center of the~algebra ${\cal A}_n$, its ``commutative ingredient'' $C^\infty(M) \otimes \{I_n\} \simeq C^\infty(M)$ where $I_n$ denotes an~``$n$ by $n$'' identity matrix.

The plan of our paper runs as follows. In Section 2, we present the~basic structure of the~model, focusing on its elements that are new as compared with previous versions of the~model. We analyse, in some detail, the~module $\textnormal{Der} \, {\cal A}_n$ of derivations of the~groupoid algebra ${\cal A}_n$ as preparation for the~study, in Section 3, of~the~geometry of the~differential algebra, understood as the~pair $({\cal A}_n$,  $\textnormal{Der} \, {\cal A}_n)$. In Section 4, we define the~Einstein--Hilbert action for our model, with the~help of which we derive the~generalized vacuum Einstein equations. Interestingly, when the~equations are projected to space--time (giving the~``general relativistic limit''), the~extra terms that appear due to our generalization, can be interpreted as ``matter terms'', not unlike in Kaluza--Klein-type models \cite{Book1,Book2,Wesson,Clifton}. We illustrate this effect in Section 5, by constructing the~simplified model with the~metric that does not ``mix'' horizontal and inner derivations (the metric matrix has a~block diagonal form). For this case, we find two ``Friedmann-like'' solutions for the~special case when $G =\mathbb{Z}_2$. Finally, in Section 6, we collect some comments and interpretative remarks.

\section{Groupoid algebra and its derivations.}
In the~following, we shall consider a~simplified version of the~groupoid unification model, in which the~Lorentz group is replaced with a~finite group $G = \{ e=g_1, g_2, g_3, \ldots, g_n \}$. This  implies that the~smooth principal bundle $E$ is trivial, $E = M \times G$. We will also assume that the~base space $M$ of the~bundle is a~compact manifold of dimension $N$.
Therefore, the~corresponding transformation groupoid is $\Gamma = E \times G = M \times G \times G$, and the~convolution in the~space $C^{\infty}(\Gamma)$ of smooth complex functions on $\Gamma$ is given by
\begin{align}
\label{conv1}
\forall \, a,b \in C^{\infty}(\Gamma) \quad (a \ast b)(x, g_k ,g_l) := \sum\limits_{m=1}^{n} a(x, g_k, g_m) b(x, g_k g_m, g_m^{-1} g_l).
\end{align}

However, by introducing the~following notation
\begin{align*}
\forall \, a \in C^{\infty}(\Gamma) \quad a_{ij}(x) := a(x, g_i, g_i^{-1} g_j)
\end{align*}
\noindent
one can easily recover from (\ref{conv1}) the~formula for matrix multiplication\footnote{Another way of realizing this fact is by considering a~suitable pair groupoid \cite{Heller2, Heller3, finite1}.}
\begin{align*}
\forall \, a,b \in C^{\infty}(\Gamma) \quad (a \ast b)_{ij}(x) := \sum\limits_{k=1}^{n} a_{ik}(x) b_{kj}(x).
\end{align*}

The convolution algebra $(C^{\infty}(\Gamma), \ast)$ can thus be identified with the~algebra ${\cal A}_n$ of ``$n$ by $n$'' matrices with entries from the~algebra $C^\infty(M)$ of smooth functions on the~manifold $M$ (with pointwise multiplication)
\begin{align*}
{\cal A}_n := \mathbb{M}_n(C^\infty(M)) = C^{\infty}(M) \otimes \mathbb{M}_n(\mathbb{C}).
\end{align*}
The center $Z({\cal A}_n)$ of this algebra consists of matrices of the~form $f \, I_n$, where $f \in C^\infty(M)$. It is therefore isomorphic to the~algebra $C^\infty(M)$.

As we briefly explained in the~introduction, our approach to noncommutative geometry of the~algebra ${\cal A}_n$ is based on the~$Z({\cal A}_n)$-module of its derivations, denoted $\textnormal{Der} \, {\cal A}_n$. Let us recall that a~\emph{derivation} of a~$\mathbb{C}$-algebra $({\cal A}_n, \ast)$ is a~$\mathbb{C}$-linear map $d: {\cal A}_n \rightarrow {\cal A}_n$ satisfying the~\emph{Leibniz rule}, namely,
\begin{align*}
\forall \, a,b \in {\cal A}_n \quad d(a \ast b) = da \ast b \, + \, a \ast db.
\end{align*}

It is straightforward to realize that the~set $\textnormal{Der} \, {\cal A}_n$ is indeed a~$Z({\cal A}_n)$-module. It also possesses the~structure of a~Lie $\mathbb{C}$-algebra with the~Lie bracket given by the~commutator
\begin{align*}
\forall \, d_1,d_2 \in \textnormal{Der} \, {\cal A}_n \quad [d_1, d_2] := d_1 \circ d_2 \, - \, d_2 \circ d_1.
\end{align*}

It is well known\footnote{In fact, one can treat this result as a~corollary of a~much more general theorem concerning derivations of tensor products of algebras \cite{TensorProds}.} that in the~case of our matrix algebra ${\cal A}_n$, the~module of its derivations can be written as a~direct sum of the~submodules of \emph{horizontal} and \emph{inner} derivations
\begin{align*}
\textnormal{Der} \, {\cal A}_n = \textnormal{Hor} \, {\cal A}_n \, \oplus \, \textnormal{Inn} \, {\cal A}_n.
\end{align*}

\emph{Horizontal derivations} are defined as liftings of the~smooth vector fields on the~manifold $M$ onto ${\cal A}_n$. More explicitly, for any smooth vector field $X \in \textnormal{Der} \, C^\infty(M)$ one can define its lifting $\bar{X} \in \textnormal{Der} \, {\cal A}_n$ as a~map acting entrywise
\begin{align*}
\forall \, a \in {\cal A}_n \quad \bar{X} a := \left( X a_{ij} \right)_{ij}.
\end{align*}

In fact, the lifting map $\bar{\ }: \textnormal{Der} \, C^\infty(M) \rightarrow \textnormal{Hor} \, {\cal A}_n$ is an~isomorphism both in the~sense of $C^\infty(M)$-modules (where from now on we shall identify the~center $Z({\cal A}_n)$ with $C^\infty(M)$) and in the~sense of Lie $\mathbb{C}$-algebras\footnote{Note that $\textnormal{Hor} \, {\cal A}_n$ possesses the~structure of a~commutative Lie algebra.}.

This means that $\textnormal{Hor} \, {\cal A}_n$ is a~locally free finitely generated $C^\infty(M)$-module, by which we mean that for any fixed chart $(U, \chi)$ on the~manifold $M$, every element $\bar{X} \in \textnormal{Hor} \, {\cal A}_n$ can be restricted (or ``localized'') to a~derivation $\bar{X}|_U \in \textnormal{Der} \, C^\infty(U \times G \times G)$, which in turn can be expressed as a~$C^\infty(U)$-linear combination of the~liftings $\overline{\tfrac{\partial}{\partial \chi^\mu}}$ of the~local vector fields induced by the~chart $\chi$. Symbolically,
\begin{align*}
\left. \left(\textnormal{Hor} \, {\cal A}_n \right) \right|_U = \textrm{span}_{C^\infty(U)} \left( \overline{\tfrac{\partial}{\partial \chi^\mu}} \right)_{\mu = 0, 1, \ldots, N-1}
\end{align*}

In the~following, we shall denote the~liftings of the~local vector fields $\overline{\tfrac{\partial}{\partial \chi^\mu}}$ simply $\partial_\mu$, suppressing both the~overline and the~reference to the~inducing chart.
\\

By an~\emph{inner derivation} induced by an~element $b \in {\cal A}_n$ one understands a~map $\textnormal{ad}_b: {\cal A}_n \rightarrow {\cal A}_n$ defined as
\begin{align*}
\forall \, a \in {\cal A}_n \quad \textnormal{ad}_b a := [b,a] = b \ast a - a \ast b.
\end{align*}

The $C^\infty(M)$-submodule $\textnormal{Inn} \, {\cal A}_n$ of all inner derivations of ${\cal A}_n$ possesses also the~structure of a~Lie algebra which, moreover, is an~ideal of the~Lie algebra $\textnormal{Der} \, {\cal A}_n$.

It is straightforward to show that the~map $\textnormal{ad}: {\cal A}_n \rightarrow \textnormal{Inn} \, {\cal A}_n$, $b \mapsto \textnormal{ad}_b$, is an~epimorphism in the~sense of $C^\infty(M)$-modules, as well as an~epimorphism going from the~Lie algebra associated\footnote{For a~given algebra $({\cal A}, \ast)$, its associated Lie algebra $({\cal A}, [\, ., . \,])$ is the~underlying vector space of ${\cal A}$ equipped with a~Lie bracket given by the~commutator $[a,b] := a \ast b - b \ast a$ for any $a,b \in {\cal A}$.} to ${\cal A}_n$ onto the~Lie algebra $\textnormal{Inn} \, {\cal A}_n$, its kernel being the~center of ${\cal A}_n$. By using the~First Isomorphism Theorem, one easily obtains an~isomorphism between the~quotient ${\cal A}_n/Z({\cal A}_n)$ (possessing both the~structure of a~$C^\infty(M)$-module and of a~Lie algebra) and $\textnormal{Inn} \, {\cal A}_n$. This quotient space, in turn, is itself isomorphic to the~space $\mathfrak{sl}_n\left(C^\infty(M)\right)$ of traceless ``$n$ by $n$'' matrices with entries from $C^\infty(M)$, again both in the~sense of $C^\infty(M)$-modules and in the~sense of Lie algebras. All in all, we can write
\begin{align*}
\textnormal{Inn} \, {\cal A}_n & = \textnormal{ad}({\cal A}_n) \cong {\cal A}_n/Z({\cal A}_n)
\\
& = \mathbb{M}_n(C^\infty(M))/ \{ f I_n \, | \, f \in C^\infty(M) \} \cong \mathfrak{sl}_n\left(C^\infty(M)\right).
\end{align*}

With a~slight abuse of notation, we shall denote the~isomorphism between $\mathfrak{sl}_n\left(C^\infty(M)\right)$ and $\textnormal{Inn} \, {\cal A}_n$ also by $\textnormal{ad}$.
With this isomorphism at hand, it is clear that $\textnormal{Inn} \, {\cal A}_n$ is a~free finitely generated $C^\infty(M)$-module, whose dimension is equal to $n^2-1$. Its basis might be determined by the~choice of the~basis in $\mathfrak{sl}_n\left(C^\infty(M)\right)$. We choose the~latter in the~following way.

Let $E_{ij}$ denote the~``$n$~by~$n$'' matrix of the~form $E_{ij} := \left( \delta_{ik} \delta_{jl} \right)_{k,l}$. Taking only those matrices $E_{ij}$ for which $k \neq l$ together with the~matrices $D_k := E_{kk}-E_{k+1, k+1}$, $(k=1,2,\ldots, n-1)$, one obtains a~convenient basis in the~$C^\infty(M)$-module $\mathfrak{sl}_n\left(C^\infty(M)\right)$.

Now, let us denote $\partial_{\{ij\}} := \textnormal{ad}_{E_{ij}}$, for $i,j = 1,2,\ldots,n$, where $i \neq j$ and $\partial_{\{k\}} := \textnormal{ad}_{D_{k}}$ for $k=1,2,\ldots, n-1$. These derivations form a~basis of the~$C^\infty(M)$-module $\textnormal{Inn} \, {\cal A}_n$. Symbolically,
\begin{align*}
\textnormal{Inn} \, {\cal A}_n = \textrm{span}_{C^\infty(M)} \left( \left( \partial_{\{ij\}} \right)_{i,j = 1, \ldots, n; \, i \neq j} \cup \left( \partial_{\{k\}} \right)_{k = 1, \ldots, n-1} \right).
\end{align*}

Summarizing, $\textnormal{Der} \, {\cal A}_n$ is thus a~locally free finitely generated $C^\infty(M)$-module; its local basis consists of the~derivations of three kinds, namely horizontal derivations and two sorts of inner derivations. To denote them we use the~same symbol ``$\partial$'' indexed according to the~following key
\begin{itemize}
\item Horizontal derivations are indexed by the~lowercase Greek letters
\\
$\mu, \nu, \lambda, \ldots$ assuming the~values $0,1,\ldots,N-1$ ($N$ stands for the~dimension of the~manifold $M$).
\item Inner derivations induced by the~matrices $E_{ij}$ are indexed by the~pairs of lowercase Latin letters in curly brackets $\{ij\}, \{kl\}, \{pq\}, \ldots$ assuming the~values $\{12\},\{13\},\ldots, \{1n\},\{21\},\{23\},\ldots,\{n,n-1\}$ (that is, out of all possible pairs only the~pairs $\{11\},\{22\},\ldots,\{nn\}$ are excluded).
\item Inner derivations induced by the~matrices $D_k$ are indexed by the~single lowercase Latin letters in curly brackets $\{a\}, \{b\}, \{c\}, \ldots$ assuming the~values $\{1\},\{2\},\ldots,\{n-1\}$.
\item Additionally, it is convenient to use the~capital Latin letters $A,B,C,\ldots$ to index \emph{all derivations} in the~basis; these indices assume all the~values listed earlier.
\item Finally, the~capital Latin letters in curly brackets $\{A\},\{B\},\{C\},\ldots$ index \emph{all inner derivations}, taking all values assumed by the~indices of the~types $\{ij\}$ and $\{a\}$
\end{itemize}

Let us illustrate the~above introduced notation to describe the~Lie algebra structure of $\textnormal{Der}\, {\cal A}_n$.
\\

\noindent
\textbf{Claim.} \ For the~elements of the~local basis of $\textnormal{Der}\, {\cal A}_n$, the~following commutation rules hold
\begin{align*}
& [\partial_{\mu}, \partial_{\nu}] = 0, \qquad [\partial_{\mu}, \partial_{\{b\}}] = 0, \qquad [\partial_{\mu}, \partial_{\{kl\}}] = 0, \qquad [\partial_{\{a\}}, \partial_{\{b\}}] = 0,
\\
& [\partial_{\{a\}}, \partial_{\{kl\}}] = \delta_{ka} \partial_{\{al\}} - \delta_{al} \partial_{\{ka\}} + \delta_{a+1, l} \partial_{\{k, a+1\}} - \delta_{k, a+1} \partial_{\{a+1, l\}},
\\
& [\partial_{\{ij\}}, \partial_{\{kl\}}] = \left\{ \begin{array}{ll}
    \partial_{\{il\}} & \textnormal{for } i \neq l \, \wedge \, k=j, \\
    -\partial_{\{kj\}} & \textnormal{for } i = l \, \wedge \, k \neq j, \\
    \partial_{\{i\}} + \partial_{\{i+1\}} + \ldots + \partial_{\{j-1\}} & \textnormal{for } i = l \, \wedge \, k = j \, \wedge \, i < j, \\
    -\left( \partial_{\{j\}} + \partial_{\{j+1\}} + \ldots + \partial_{\{i-1\}} \right) & \textnormal{for } i = l \, \wedge \, k = j \, \wedge \, i > j,\\
    0 & \textnormal{for } i \neq l \, \wedge \, k \neq j.
  \end{array}\right.
\end{align*}
The claim is a~straightforward consequence of the~following commutation rule for the~matrices $E_{ij}$:
\begin{align*}
E_{ij} E_{kl} - E_{kl} E_{ij} = \delta_{kj} E_{il} - \delta_{il} E_{kj}.
\end{align*}

In other words, because of the~noncommutativity of ${\cal A}_n$, some of the~structure constants\footnote{The structure constants $\textbf{c}_{AB}^{\ \ \ \, C}$, here interpreted as constant functions on $M$, are defined by $[\partial_A, \partial_B] = \textbf{c}_{AB}^{\ \ \ \, C} \partial_C$.} $\textbf{c}_{AB}^{\ \ \ \, C}$ of the~Lie algebra $\textnormal{Der}\, {\cal A}_n$ are nontrivial. To be precise,
\begin{align}
\label{sc1}
& \textbf{c}_{\{a\} \{kl\}}^{\quad \quad \{pq\}} = \delta_{pa} \delta_{ql} \delta_{ka} - \delta_{pk} \delta_{qa} \delta_{al} + \delta_{pk} \delta_{q,a+1} \delta_{a+1,l} - \delta_{p,a+1} \delta_{q,l}  \delta_{k,a+1},
\\
\label{sc2}
& \textbf{c}_{\{ij\} \{b\} }^{\quad \quad \{pq\}} = - \textbf{c}_{\{b\} \{ij\}}^{\quad \quad \{pq\}},
\\
\label{sc3}
& \textbf{c}_{\{ij\} \{kl\}}^{\quad \quad \ \{pq\}} = \delta_{pi} \delta_{ql} \delta_{kj} - \delta_{pk} \delta_{qj} \delta_{il},
\\
\label{sc4}
& \textbf{c}_{\{ij\} \{kl\}}^{\quad \quad \ \{c\}} = \left\{ \begin{array}{ll}
    \delta_{il} \delta_{kj} \left( \delta_{ci} + \delta_{c,i+1} + \ldots + \delta_{c,j-1} \right) & \textnormal{for } i < j,
    \\
    - \delta_{il} \delta_{kj} \left( \delta_{cj} + \delta_{c,j+1} + \ldots + \delta_{c,i-1} \right) & \textnormal{for } i > j,
    \end{array} \right.
\end{align}
\noindent
and all other structure constants are zero.
\\

To further get used to the~notation, let us consider an~example in which $G = \mathbb{Z}_2$. The~algebra we study is then ${\cal A}_2 := \mathbb{M}_2(C^\infty(M)) = C^{\infty}(M) \otimes \mathbb{M}_2(\mathbb{C})$.
In this case, the~$C^\infty(M)$-module $\textnormal{Inn} \, {\cal A}_2$ is spanned by three derivations: $\partial_{\{12\}}, \partial_{\{21\}}$ and $\partial_{\{1\}}$, that are induced by the~following matrices, respectively:
\begin{align*}
E_{12} = \left[\begin{array}{cc}
    0 & 1 \\
    0 & 0
  \end{array}\right],
\qquad
E_{21} = \left[\begin{array}{cc}
    0 & 0 \\
    1 & 0
  \end{array}\right],
\qquad
D_{1} = \left[\begin{array}{cc}
    1 & 0 \\
    0 & -1
  \end{array}\right].
\end{align*}

One has the~following commutation relations:
\begin{align*}
[ \partial_{\{1\}}, \partial_{\{12\}} ] = 2 \partial_{\{12\}}, \qquad [ \partial_{\{1\}}, \partial_{\{21\}} ] = -2 \partial_{\{21\}}, \qquad [ \partial_{\{12\}}, \partial_{\{21\}} ] = \partial_{\{1\}}.
\end{align*}
These relations translate into the~following values of the~nonzero structure constants:
\begin{align*}
&\textbf{c}_{\{1\}\{12\}}^{\qquad \ \{12\}} = - \textbf{c}_{\{12\}\{1\}}^{\qquad \ \{12\}} = 2,
\\
&\textbf{c}_{\{1\}\{21\}}^{\qquad \ \{21\}} = - \textbf{c}_{\{21\}\{1\}}^{\qquad \ \{21\}} = -2,
\\
&\textbf{c}_{\{12\}\{21\}}^{\qquad \ \ \{1\}} = - \textbf{c}_{\{21\}\{12\}}^{\qquad \ \ \{1\}} = 1.
\end{align*}

\section{Geometry of the~differential algebra \\ $( {\cal A}_n, \textnormal{Der} \, {\cal A}_n)$.}
In this section,  we construct the~elements of the~pseudo-Riemannian geometry of the~algebra ${\cal A}_n$. We proceed along the~lines first proposed by Parfionov and Zapatrin \cite{Parfionov} and further developed in \cite{Heller2,Structured,Sasin1}. Our ``toy version'' of the~model, albeit simple, still offers a~nontrivial extension of the~standard pseudo-Riemannian geometry.

For brevity, let us denote the~$C^\infty(M)$-module $\textnormal{Der}({\cal A}_n)$ by $V$. We define $V^\ast \equiv \textnormal{Hom}_{C^\infty(M)}\left(V, C^\infty(M)\right)$ to be its dual.

Let now ${\cal G}: V \times V \rightarrow C^\infty(M)$ be a~symmetric $C^\infty(M)$-bilinear map, the~\textit{metric} in our model. We assume that ${\cal G}$ is nondegenerate, that is, that the~map $\Phi_{{\cal G}}: V \rightarrow V^\ast$, given by
\begin{align*}
\forall \, u,v \in V \quad \Phi_{{\cal G}}(u)(v)={\cal G}(u,v),
\end{align*}
\noindent
is an~isomorphism of $C^\infty(M)$-modules.

For the~sake of further calculations, it is convenient to express the~metric in the~local basis of $V$ introduced in the~previous section. We thus have the~\emph{metric matrix} $(g_{AB})$ with the~entries
\begin{align*}
g_{AB} := {\cal G} ( \partial_A, \partial_B ).
\end{align*}
It is a~square, symmetric and nonsingular matrix of order $N + n^2 - 1$. We denote its inverse matrix by $(g^{AB})$. Exactly as in the~standard case, the~metric matrix and its inverse can be used to lower or raise indices. However, one should remember that this concerns only the~indices of the~type $A, B, C, \ldots$\footnote{Unless the~metric matrix is block diagonal, see Section 5.}.

For instance, we shall be using the~\emph{completely covariant structure constants} $\textbf{c}_{ABC} := \textbf{c}_{AB}^{\ \ \ \, D}g_{DC}$. For our algebra, they can be expressed as
\begin{align}
\label{ccsc1}
& \textbf{c}_{\{a\} \{kl\} C} = \delta_{ka} g_{\{al\} C} - \delta_{al} g_{\{ka\} C} + \delta_{a+1,l} g_{\{k,a+1\} C} - \delta_{k,a+1} g_{\{a+1,l\} C},
\\
\label{ccsc2}
& \textbf{c}_{\{ij\} \{b\} C} = -\textbf{c}_{\{b\} \{ij\} C},
\\
\label{ccsc3}
& \textbf{c}_{\{ij\} \{kl\} C} = \left\{ \begin{array}{ll}
    \delta_{il} \delta_{kj} \left( g_{\{i\} C} + g_{\{i+1\} C} + \ldots + g_{\{j-1\} C} \right) \\ + \delta_{kj} g_{\{il\} C} - \delta_{il} g_{\{kj\} C} &  \textnormal{for } i < j,
    \\
    - \delta_{il} \delta_{kj} \left( g_{\{j\} C} + g_{\{i+1\} C} + \ldots + g_{\{i-1\} C} \right) \\ + \delta_{kj} g_{\{il\} C} - \delta_{il} g_{\{kj\} C} & \textnormal{for } i > j
    \end{array} \right.
\end{align}
\noindent
with all the~remaining ones equal to zero.

We are now ready to define the~\emph{Levi-Civita preconnection} $\nabla^\ast: V \times V \rightarrow V^\ast$ by using the~Koszul formula
\begin{align}
\begin{split}
\label{Koszul}
\left(\nabla^\ast_u v \right)(w) & := \tfrac{1}{2} \big[ u\left({\cal G} (v,w)\right) + v\left({\cal G} (u,w)\right) - w\left({\cal G} (u,v)\right) \big.
\\
& \ + \big. {\cal G}(w,[u,v]) + {\cal G}(v,[w,u]) - {\cal G}(u, [v,w]) \big]
\end{split}
\end{align}
\noindent
for any $u,v,w \in V$.

The components $\Gamma_{ABC}$ of the~Levi-Civita preconnection in the~local basis (the~Christoffel symbols of the~first kind) are expressed by the~following general formula:
\begin{align}
\begin{split}
\label{preconnection_components}
\Gamma_{ABC} & := \left(\nabla^\ast_{\partial_C} \partial_B \right)(\partial_A)
\\
& \, = \tfrac{1}{2}\left( \partial_C g_{AB} + \partial_B g_{AC} - \partial_A g_{BC} + \textbf{c}_{CBA} + \textbf{c}_{ACB} - \textbf{c}_{BAC} \right).
\end{split}
\end{align}

Let us notice that, unlike in the~standard case, the~components $\Gamma_{ABC}$ are in general not symmetric with respect to interchanging the~second and third indices,
\begin{align*}
\Gamma_{ABC} - \Gamma_{ACB} = \textbf{c}_{CBA}.
\end{align*}

The \emph{Levi-Civita connection} $\nabla: V \times V \rightarrow V$ is defined as
\begin{align*}
\nabla := \Phi_{{\cal G}}^{-1} \circ \nabla^{\ast}.
\end{align*}
Its components in the~local basis (the~Christoffel symbols of the~second kind) are exactly the~components of the~Levi-Civita preconnection with the~first index raised $\Gamma^A_{\ BC}$. This means that they can be equivalently defined by the~equality
\begin{align}
\label{connection_components_def}
\nabla_{\partial_C} \partial_B = \Gamma^A_{\ BC} \partial_A
\end{align}
\noindent
and that they can be expressed as
\begin{align}
\label{connection_components}
\Gamma^A_{\ BC} = \tfrac{1}{2}g^{AD} \left( \partial_C g_{DB} + \partial_B g_{DC} - \partial_D g_{BC} + \textbf{c}_{CBD} + \textbf{c}_{DCB} - \textbf{c}_{BDC} \right).
\end{align}

Just as for the~preconnection components, we have the~asymmetry
\begin{align}
\label{asymetria2}
\Gamma^A_{\ BC} - \Gamma^A_{\ CB} = \textbf{c}^{\ \ \ \, A}_{CB}.
\end{align}

The Levi-Civita connection enjoys all the~properties of its standard counterpart, that is\footnote{Properties $1^\circ$--$6^\circ$ follow directly from the Koszul formula (\ref{Koszul}).}
\begin{align*}
& 1^\circ \quad \nabla_{u + v} w = \nabla_{u} w + \nabla_{v} w,
\\
& 2^\circ \quad \nabla_{f u} v = f \, \nabla_u v,
\\
& 3^\circ \quad \nabla_u (v + w) = \nabla_u v + \nabla_u w,
\\
& 4^\circ \quad \nabla_u (f v) = u(f) v + f \, \nabla_u v,
\\
& 5^\circ \quad \nabla_u v - \nabla_v u - [u, v] = 0
\\
& \qquad (\textit{torsion-freeness}),
\\
& 6^\circ \quad w\left({\cal G} (u,v)\right) = {\cal G} \left( \nabla_w u , v \right) + {\cal G} \left( u, \nabla_w v \right)
\\
& \qquad (\textit{metric compatibility})
\end{align*}
\noindent
for all $u, v, w \in V$ and $f \in C^\infty(M)$. Moreover, just like in the~standard case, the~Levi-Civita connection is the~unique map $V \times V \rightarrow V$ that satisfies $5^\circ$ and $6^\circ$.

With the~help of the~Levi-Civita connection, one can introduce the~curvature tensors. The~\emph{Riemann curvature tensor} $R: V \times V \times V \rightarrow V$, $(u,v,w) \mapsto R(u,v)w$ is defined as
\begin{align*}
R(u,v)w := \nabla_u \nabla_v w - \nabla_v \nabla_u w - \nabla_{[u,v]}w.
\end{align*}
Its components $R^C_{\ DAB}$ in the~local basis of $V$, defined by the~formula
\begin{align*}
R(\partial_A, \partial_B) \partial_D = R^C_{\ DAB} \partial_C,
\end{align*}
\noindent
can be expressed as
\begin{align}
\label{riemann2}
R^C_{\ DAB} = \partial_A \Gamma^C_{\ DB} - \partial_B \Gamma^C_{\ DA} + \Gamma^K_{\ DB} \Gamma^C_{\ KA} - \Gamma^K_{\ DA} \Gamma^C_{\ KB} - \textbf{c}_{AB}^{\ \ \ \, K} \Gamma^C_{\ DK}.
\end{align}

The map $R$ enjoys the~usual Riemann tensor symmetries which, when expressed in the~local basis, read
\begin{align}
\begin{split}
\label{Rsymmetries}
& R_{CDAB} = - R_{DCAB} = - R_{CDBA} = R_{ABCD},
\\
& R^C_{\ DAB} + R^C_{\ BDA} + R^C_{\ ABD} = 0.
\end{split}
\end{align}

Thanks to the~finite-dimensionality of $V$, one can use the~standard tensor contraction operation to define the~Ricci tensor, $\textbf{ric}: V \times V \rightarrow C^\infty(M)$, and the~curvature scalar $r \in C^\infty(M)$.
The components of the~Ricci tensor in the~local basis read
\begin{align}
\label{ricci_def}
\textbf{ric}_{AB} := R^C_{\ ACB},
\end{align}
\noindent
and the~curvature scalar is
\begin{align}
\label{scalar_def}
r := g^{AB} \textbf{ric}_{AB} = g^{AB} R^C_{\ ACB}.
\end{align}
Note that $\textbf{ric}$ is a~symmetric tensor, $\textbf{ric}_{AB} = \textbf{ric}_{BA}$.

Having defined all the~basic pseudo-Riemannian-geometric elements of the~``toy version'' of our model, we are ready to investigate some of its relativistic aspects.

\section{Generalized Einstein equations from the~action principle.}
In this Section, after providing a~straightforward generalization of the~Einstein--Hilbert action, we derive the~generalized vacuum Einstein equations.

Let us consider the~following Einstein--Hilbert action functional
\begin{align}
\label{action}
S_{EH} := \int r \sqrt{|g|} \, d^N x,
\end{align}
\noindent
where again $N = \dim M$ and $g$ denotes the~determinant of the~metric matrix $(g_{AB})$. We postulate no additional matter term.

Let us vary $S_{EH}$ with respect to $\delta g^{A B}$
\begin{align}
\begin{split}
\label{variation1}
\delta S_{EH} & = \int \left( \textbf{ric}_{A B} - \frac{1}{2} r g_{A B} \right) \delta g^{A B} \sqrt{|g|} \, d^Nx
\\
& \quad + \int \delta \textbf{ric}_{A B} g^{A B} \sqrt{|g|} \, d^Nx.
\end{split}
\end{align}

We will now show that the~rightmost integral in (\ref{variation1}) can be omitted, because its integrand can be expressed as a~divergence, namely
\begin{align}
\label{variation2}
\delta \textbf{ric}_{A B} g^{A B} \sqrt{|g|} = \partial_\mu \left[ \left(g^{A B} \delta \Gamma^\mu_{\ A B} - g^{\mu B} \delta \Gamma^{A}_{\ B A}\right) \sqrt{|g|} \right].
\end{align}

We begin the~proof of this statement by making the~following observations that will greatly simplify further computations.
\begin{enumerate}[(i)]
\item The~structure constants $\textbf{c}^{\ \ \ \, C}_{AB}$ do not depend on the~components of the~metric matrix $g_{AB}$, and hence their variations vanish
\begin{align*}
\delta \textbf{c}^{\ \ \ \, C}_{AB} = 0.
\end{align*}
\item The~structure constants $\textbf{c}^{\ \ \ \, C}_{AB}$ do not depend on the~space--time coordinates, and hence they vanish under the~action of \emph{any} derivation
\begin{align*}
\partial_D \textbf{c}^{\ \ \ \, C}_{AB} = 0.
\end{align*}
\item Any contraction of $\textbf{c}^{\ \ \ \, C}_{AB}$ yields zero
\begin{align*}
\textbf{c}^{\ \ \ \, A}_{AB} = 0, \qquad \textbf{c}^{\ \ \ \, B}_{AB} = 0, \qquad g^{AB}\textbf{c}^{\ \ \ \, C}_{AB} = 0.
\end{align*}
\item Although the~Christoffel symbols are in general asymmetric with respect to interchanging the~second and third indices (\ref{asymetria2})
\begin{align*}
\Gamma^A_{\ BC} - \Gamma^A_{\ CB} = \textbf{c}^{\ \ \ \, A}_{CB},
\end{align*}
\noindent
the symmetry is recovered for contractions, derivatives and variations of the~symbols
\begin{align*}
\Gamma^A_{\ AC} = \Gamma^A_{\ CA}, \qquad \partial_D \Gamma^A_{\ BC} = \partial_D \Gamma^A_{\ CB}, \qquad \delta \Gamma^A_{\ BC} = \delta \Gamma^A_{\ CB}.
\end{align*}

Moreover,
\begin{align*}
\Gamma^A_{\ AC} = \tfrac{1}{2} g^{AB} \partial_C g_{AB} = - \tfrac{1}{2} g_{AB} \partial_C g^{AB}.
\end{align*}
\end{enumerate}

To prove (iii) let us directly compute the~contraction $\textbf{c}^{\ \ \ \, B}_{AB}$, and let us do this separately for all possible kinds of the~index $A$. By (\ref{sc1}) we obtain
\begin{align*}
\textbf{c}^{\ \ \ \, B}_{\mu B} & = 0 \quad \textnormal{(trivially)},
\\
\textbf{c}^{\quad \ \ \, B}_{\{a\} B} & = \textbf{c}^{\quad \quad \{ij\}}_{\{a\} \{ij\}}
\\
& = \sum\limits_{i,j=1}^n \left( \delta_{ia} \delta_{jj} \delta_{ia} - \delta_{ii} \delta_{ja} \delta_{aj} + \delta_{ii} \delta_{j,a+1} \delta_{a+1,j} - \delta_{i,a+1} \delta_{jj} \delta_{i,a+1} \right)
\\
& = n-n+n-n = 0,
\\
\textbf{c}^{\quad \ \ \ B}_{\{kl\} B} & = \textbf{c}^{\quad \quad \ \{ij\}}_{\{kl\} \{ij\}} =
\sum\limits_{i,j=1}^n \left( \delta_{ik} \delta_{jj} \delta_{il} - \delta_{ii} \delta_{jl} \delta_{kj} \right) = n \delta_{kl} - n \delta_{kl} = 0.
\end{align*}

The other two contractions vanish on the~strength of the~skew-symmetry $\textbf{c}^{\ \ \ \, C}_{AB} = - \textbf{c}^{\ \ \ \, C}_{BA}$.

As for (iv), it is a~straightforward consequence of (i-iii) and (\ref{asymetria2}).

We are now ready to prove (\ref{variation2}). By expanding the~expression $\delta \textbf{ric}_{A B} g^{A B}$, we obtain
\begin{align}
\label{variation3}
\begin{split}
\delta \textbf{ric}_{A B} g^{A B} & = \delta R^C_{\ ACB} g^{A B}
\\
& = g^{AB} \left( \partial_C \delta \Gamma^C_{\ AB} - \partial_B \delta \Gamma^C_{\ AC} + \delta \Gamma^K_{\ AB} \Gamma^C_{\ KC} + \Gamma^K_{\ AB} \delta \Gamma^C_{\ KC} \right.
\\
& \quad \quad \quad \ \left. - \, \delta \Gamma^K_{\ AC} \Gamma^C_{\ KB} - \Gamma^K_{\ AC} \delta \Gamma^C_{\ KB} - \textbf{c}^{\ \ \ \, K}_{CB} \delta \Gamma^C_{\ AK} \right),
\end{split}
\end{align}
\noindent
where we have used (i) and the~fact that the~derivatives of variations are equal to the~variations of derivatives.

The last three terms in (\ref{variation3}) can be merged into a~single term with the~help of (\ref{asymetria2}) and (iv). Concretely, one can show that
\begin{align*}
& g^{AB} \left( \delta \Gamma^K_{\ AC} \Gamma^C_{\ KB} + \Gamma^K_{\ AC} \delta \Gamma^C_{\ KB} + \textbf{c}^{\ \ \ \, K}_{CB} \delta \Gamma^C_{\ AK} \right) = 2 g^{AB} \Gamma^K_{\ BC} \delta \Gamma^C_{\ AK}.
\end{align*}
This term, in turn, can be expressed as
\begin{align*}
2 g^{AB} \Gamma^K_{\ BC} \delta \Gamma^C_{\ AK} = - \partial_C g^{KA} \delta \Gamma^C_{\ AK}.
\end{align*}
Indeed, by expanding the~Christoffel symbol $\Gamma^K_{\ BC}$ one gets
\begin{align*}
& 2 g^{AB} \Gamma^K_{\ BC} \delta \Gamma^C_{\ AK}
\\
& = g^{AB} g^{KD} \big( \partial_C g_{DB} + \partial_B g_{DC} - \partial_D g_{BC} + \textbf{c}_{CBD} + \textbf{c}_{DCB} - \textbf{c}_{BDC} \big) \delta \Gamma^C_{\ AK}
\\
& = g^{AB} g^{KD} \partial_C g_{DB} \delta \Gamma^C_{\ AK}
\\
& \quad + g^{AB} g^{KD} \big( \underbrace{\partial_B g_{DC} - \partial_D g_{BC}}_{= \, 0} + \underbrace{\textbf{c}_{CBD} - \textbf{c}_{CDB}}_{= \, 0} - \underbrace{\textbf{c}_{BDC}}_{= \, 0} \big) \delta \Gamma^C_{\ AK}
\\
& = - \partial_C g^{KA} \delta \Gamma^C_{\ AK},
\end{align*}
\noindent
where all the~underbraced terms vanish because the~term $g^{AB} g^{KD} \delta \Gamma^C_{\ AK}$ is symmetric with respect to interchanging the~indices $B$ and $D$. Moreover, in the~last equality we have employed the~well-known formula for the~derivative of the~matrix inverse.

Thus, equality (\ref{variation3}) can be rewritten in the~form
\begin{align}
\begin{split}
\label{variation4}
\delta \textbf{ric}_{A B} g^{A B} & = g^{AB} \partial_C \delta \Gamma^C_{\ AB} - g^{AB} \partial_B \delta \Gamma^C_{\ AC} + g^{AB} \delta \Gamma^K_{\ AB} \Gamma^C_{\ KC}
\\
& \quad + g^{AB} \Gamma^K_{\ AB} \delta \Gamma^C_{\ KC} + \partial_C g^{KA} \delta \Gamma^C_{\ AK}.
\end{split}
\end{align}

Let us now expand the~expression $\frac{1}{\sqrt{|g|}} \partial_C \left[ \left(g^{A B} \delta \Gamma^C_{\ A B} - g^{C B} \delta \Gamma^{A}_{\ B A}\right) \sqrt{|g|} \right]$. By using the~fact that, by (iv),
\begin{align*}
\partial_C \sqrt{|g|} = \tfrac{1}{2} \sqrt{|g|} g^{DK} \partial_C g_{DK} = \sqrt{|g|} \Gamma^K_{\ KC},
\end{align*}
\noindent
one has
\begin{align}
\begin{split}
\label{variation5}
& \frac{1}{\sqrt{|g|}} \partial_C \left[ \left(g^{A B} \delta \Gamma^C_{\ A B} - g^{C B} \delta \Gamma^{A}_{\ B A}\right) \sqrt{|g|} \right]
\\
& = \partial_C g^{AB} \delta \Gamma^C_{\ A B} + g^{AB} \partial_C \delta \Gamma^C_{\ A B} - \partial_C g^{C B} \delta \Gamma^{A}_{\ B A} - g^{C B} \partial_C \delta \Gamma^{A}_{\ B A}
\\
& \quad + \Gamma^K_{\ KC} g^{A B} \delta \Gamma^C_{\ A B} - \Gamma^K_{\ KC} g^{C B} \delta \Gamma^{A}_{\ B A}.
\end{split}
\end{align}

Comparing (\ref{variation5}) with (\ref{variation4}) (and renaming some of the~dummy indices), one concludes that they are in fact equal if and only if
\begin{align}
\label{variation6}
& \left( \Gamma^K_{\ KC} g^{CB} + \Gamma^B_{\ CK} g^{CK} + \partial_K g^{KB} \right ) \delta \Gamma^A_{\ AB} = 0.
\end{align}
But this is indeed the~case because, by direct calculation one gets
\begin{align*}
& \Gamma^K_{\ KC} g^{CB} + \Gamma^B_{\ CK} g^{CK}
\\
& = \tfrac{1}{2} g^{CB} g^{DE} \partial_C g_{DE}
\\
& \quad + \tfrac{1}{2} g^{CK} g^{BD} \left( \partial_K g_{DC} + \partial_C g_{DK} - \partial_D g_{CK} + \textbf{c}_{KCD} + \textbf{c}_{DKC} - \textbf{c}_{CDK} \right)
\\
& = \tfrac{1}{2} g^{CB} g^{DE} \partial_C g_{DE} + g^{CK} g^{BD} \partial_K g_{DC} - \tfrac{1}{2} g^{CK} g^{BD} \partial_D g_{CK}
\\
& = g^{CK} g^{BD} \partial_K g_{DC} = - \partial_K g^{BK},
\end{align*}
\noindent
where all the~terms involving structure constants vanish by (iii), and in the~last line we again use the~formula for the~derivative of the~matrix inverse. We have thus proven (\ref{variation6}) and obtained the~following equality:
\begin{align}
\label{variation7}
\delta \textbf{ric}_{A B} g^{A B} = \frac{1}{\sqrt{|g|}} \partial_C \left[ \left(g^{A B} \delta \Gamma^C_{\ A B} - g^{C B} \delta \Gamma^{A}_{\ B A}\right) \sqrt{|g|} \right].
\end{align}
But this equality immediately implies (\ref{variation2}) because one can replace the~index $C$ on the~right-hand side of (\ref{variation7}) with the~index $\mu$ (this can be done because every inner derivation yields zero when acting on any element of the~center).

Having proven (\ref{variation2}), we can omit the~second integral in (\ref{variation1}) and use the~action principle to obtain the~following generalized Einstein equations:
\begin{align*}
\textbf{ric}_{A B} - \frac{1}{2} r g_{A B} = 0
\end{align*}
\noindent
which can be immediately reduced to
\begin{align}
\label{Einstein1}
\textbf{ric}_{A B} = 0.
\end{align}

Therefore, one can say that endowing the~differential algebra $\left( {\cal A}_n, \textnormal{Der} \, {\cal A}_n \right)$ with the~Einstein--Hilbert action given by (\ref{action}) grants it the~structure of a~noncommutative Einstein algebra \cite{Geroch,Heller1a}\footnote{Technically, one has yet to impose the~condition that the~metric matrix is lorentzian and that the~Levi-Civita connection $\nabla$ can be used to define the~covariant derivative. Both these requirements can be here easily met.}.

Interestingly, these generalized Einstein equations, after being projected onto space--time $M$, have a~richer form than the~standard Einstein equations. This is because of the~extra terms coming from additional components of the~metric. These extra terms could be interpreted as an~``$N$-dimensional matter--energy'' induced by the~generalized (vacuum) Einstein equations (similarly as in the~Kaluza--Klein-type theories \cite{Book1,Book2,Wesson,Clifton}). We shall demonstrate this effect by considering an~example of a~simple block diagonal metric.

\section{Model with a simple metric.}

From now on, we consider a~metric ${\cal G}$ that does not ``mix'' the~horizontal derivations with the~inner derivations
\begin{align*}
\forall \, \bar{X} \in \textnormal{Hor} \, {\cal A}_n \ \ \forall \, \textnormal{ad}_a \in \textnormal{Inn} \, {\cal A}_n \ \quad {\cal G}\left( \bar{X}, \textnormal{ad}_a \right) = {\cal G}\left( \textnormal{ad}_a, \bar{X} \right) = 0.
\end{align*}

Therefore, the~metric matrix $(g_{A B})$ assumes the~following block diagonal form
\begin{align}
\label{block_metric}
g_{A B} =
\left[\begin{array}{cc}
    g_{\mu \nu} & 0 \\
    0 & g_{\{A\} \{B\}}
  \end{array}\right].
\end{align}
\noindent
We recall that $(g_{\mu \nu})$ is an~``$N$~by~$N$'' matrix ($N = \dim M$) and $(g_{\{A\} \{B\}})$ is a~``$\left(n^2-1\right)$ by $\left(n^2-1\right)$'' ($n = |G|$) matrix.
Of course, the~inverse metric matrix $(g^{A B})$ is also block diagonal.

Let us emphasize that by restricting our attention to the~metric matrices of the~form (\ref{block_metric}) one does not erase the~noncommutativity of the~model as the~structure constants $\textbf{c}_{AB}^{\ \ \ \, C}$ are still given by formulae (\ref{sc1}--\ref{sc4}). On the~other hand, some more of the~completely covariant structure constants are now vanishing. In fact, formulae (\ref{ccsc1}--\ref{ccsc3}) imply that
 \begin{enumerate}
\item[(v)] Only the~completely covariant structure constants of the~form $\textbf{c}_{\{A\}\{B\}\{C\}}$ are nonzero.
\end{enumerate}

With properties (i--iv) (listed in Section 4) and property (v), we are ready to compute the~Christoffel symbols and the~curvature tensor components associated with the~block diagonal metric matrix (\ref{block_metric}).

By using general formula (\ref{preconnection_components}) for the~components $\Gamma_{ABC}$ of the~Levi-Civita preconnection one obtains
\begin{align}
\label{preconnection_compo1}
\Gamma_{\lambda \mu \nu} & = \tfrac{1}{2}\left( \partial_\nu g_{\lambda \mu} + \partial_\mu g_{\lambda \nu} - \partial_\lambda g_{\mu \nu} \right) = \widetilde{\Gamma}_{\lambda \mu \nu},
\\
\label{preconnection_compo2}
\Gamma_{\{A\} \mu \nu} & = 0, \quad \ \Gamma_{\lambda \{B\} \nu} = 0, \quad \ \Gamma_{\lambda \mu \{C\}} = 0,
\\
\label{preconnection_compo3}
\Gamma_{\lambda \{B\} \{C\}} & = - \tfrac{1}{2} \partial_\lambda g_{\{B\} \{C\}},
\\
\label{preconnection_compo4}
\Gamma_{\{A\} \mu \{C\}} & = \tfrac{1}{2} \partial_\mu g_{\{A\} \{C\}},
\\
\label{preconnection_compo5}
\Gamma_{\{A\} \{B\} \nu} & = \tfrac{1}{2} \partial_\nu g_{\{A\} \{B\}},
\\
\label{preconnection_compo6}
\Gamma_{\{A\} \{B\} \{C\}} & = \tfrac{1}{2} \left( \textbf{c}_{\{C\}\{B\}\{A\}} + \textbf{c}_{\{A\}\{C\}\{B\}} + \textbf{c}_{\{A\}\{B\}\{C\}} \right),
\end{align}
\noindent
where from now on a~tilde ($\widetilde{\ }$) above an~object will mean that the~object is ``classical'', i.e. it is obtained and used according to the~standard, pseudo-Riemannian-geometric formulae that employ only the~``horizontal'' part of the~metric matrix $(g_{\mu \nu})$.

By raising the~first index in (\ref{preconnection_compo1}--\ref{preconnection_compo6}) one immediately obtains the~components $\Gamma^A_{\ BC}$ of the~Levi-Civita connection
\begin{align}
\label{connection_compo1}
\Gamma^\kappa_{\ \mu \nu} & = \tfrac{1}{2} g^{\kappa \lambda} \left( \partial_\nu g_{\lambda \mu} + \partial_\mu g_{\lambda \nu} - \partial_\lambda g_{\mu \nu} \right) = \widetilde{\Gamma}^\kappa_{\ \mu \nu},
\\
\label{connection_compo2}
\Gamma^{\{A\}}_{\quad \mu \nu} & = 0, \quad \ \Gamma^\kappa_{\ \{B\} \nu} = 0, \quad \ \Gamma^\kappa_{\ \mu \{C\}} = 0,
\\
\label{connection_compo3}
\Gamma^\kappa_{\ \{B\} \{C\}} & = - \tfrac{1}{2} g^{\kappa \lambda} \partial_\lambda g_{\{B\} \{C\}},
\\
\label{connection_compo4}
\Gamma^{\{A\}}_{\quad \mu \{C\}} & = \tfrac{1}{2} g^{\{A\} \{D\}} \partial_\mu g_{\{D\} \{C\}},
\\
\label{connection_compo5}
\Gamma^{\{A\}}_{\quad \{B\} \nu} & = \tfrac{1}{2} g^{\{A\}\{D\}} \partial_\nu g_{\{D\} \{B\}},
\\
\label{connection_compo6}
\Gamma^{\{A\}}_{\quad \{B\} \{C\}} & = \tfrac{1}{2} g^{\{A\}\{D\}} \left( \textbf{c}_{\{C\}\{B\}\{D\}} + \textbf{c}_{\{D\}\{C\}\{B\}} + \textbf{c}_{\{D\}\{B\}\{C\}} \right).
\end{align}

Thanks to the~block-diagonality of the~metric matrix, these formulae are still relatively simple.

Let us take a~closer look at (\ref{connection_compo1}) and (\ref{connection_compo2}). By (\ref{connection_components_def}) they imply that
\begin{align}
\label{connection_extension}
\forall \, \bar{X}, \bar{Y} \in \textnormal{Hor} \, {\cal A}_n \quad \nabla_{\bar{X}} \bar{Y} = \overline{\widetilde{\nabla}_X Y}.
\end{align}

In other words, for the~metrics whose matrices are of the~form (\ref{block_metric}), the~Levi-Civita connection acts on horizontal derivations in exactly the~same way as does its classical counterpart. One can thus say that $\nabla$ \emph{is a~(nontrivial) extension of}~$\widetilde{\nabla}$.

For comparison as well as for further use, let us introduce the~\emph{trivial extension of the~classical Levi-Civita connection} $\widehat{\nabla}: V \times V \rightarrow V$ via
\begin{align}
\label{trivial_extension}
\forall \, \bar{X}, \bar{Y} \in \textnormal{Hor} \, {\cal A}_n \quad \forall \, \textnormal{ad}_a, \textnormal{ad}_b \in \textnormal{Inn} \, {\cal A}_n \quad
\widehat{\nabla}_{\bar{X} + \textnormal{ad}_a} \left( \bar{Y} + \textnormal{ad}_b \right) := \overline{\widetilde{\nabla}_X Y}.
\end{align}
It is indeed a~well-defined connection (i.e. it has the~properties $1^\circ$--$4^\circ$ listed in Section 3.), however it is neither torsion-free nor metric-compatible (i.e. it fails to satisfy $5^\circ$ and $6^\circ$).

We now move to computing the~components of the~Riemann tensor. Either directly from (\ref{connection_extension}) or through computations employing (\ref{riemann2}) one obtains
\begin{align}
\label{riemann_compo1}
R^\rho_{\ \sigma \mu \nu} & = \partial_\mu \Gamma^\rho_{\ \sigma \nu} - \partial_\nu \Gamma^\rho_{\ \sigma \mu} + \Gamma^\kappa_{\ \sigma \nu} \Gamma^\rho_{\ \kappa \mu} - \Gamma^\kappa_{\ \sigma \mu} \Gamma^\rho_{\ \kappa \nu} = \widetilde{R}^\rho_{\ \sigma \mu \nu},
\\
\label{riemann_compo2}
R^{\{C\}}_{\quad \sigma \mu \nu} & = 0, \quad \ R^\rho_{\ \{D\} \mu \nu} = 0, \quad \ R^\rho_{\ \sigma \{A\} \nu} = 0, \quad \ R^\rho_{\ \sigma \mu \{B\}} = 0.
\end{align}

By using (\ref{riemann2}), one also obtains that
\begin{align}
\begin{split}
\label{riemann_compo3}
R^{\{C\}}_{\quad \sigma \{A\} \nu} & = -\tfrac{1}{2} g^{\{C\} \{E\}} \partial_\nu \partial_\sigma g_{\{E\} \{A\}} + \tfrac{1}{2} \Gamma^\lambda_{\ \sigma \nu} g^{\{C\} \{E\}} \partial_\lambda g_{\{E\} \{A\}}
\\
& \quad - \tfrac{1}{4} \partial_\nu g^{\{C\} \{E\}} \partial_\sigma g_{\{E\} \{A\}}
\\
& = -\tfrac{1}{2} g^{\{C\} \{E\}} \widehat{\nabla}_\nu \partial_\sigma g_{\{E\} \{A\}} - \tfrac{1}{4} \partial_\nu g^{\{C\} \{E\}} \partial_\sigma g_{\{E\} \{A\}},
\end{split}
\end{align}
\noindent
where $\widehat{\nabla}$ denotes the~covariant derivative arising from the~trivial extension of the~classical Levi-Civita connection (\ref{trivial_extension}). For practical purposes it is important to notice that when acting on tensors written in the~index notation, $\widehat{\nabla}_\nu$ takes into account only their lowercase Greek indices. That is why in (\ref{riemann_compo3}) one could use the~fact that $\widehat{\nabla}_\nu \partial_\sigma g_{\{E\} \{A\}} = \partial_\nu \partial_\sigma g_{\{E\} \{A\}} - \Gamma^\lambda_{\ \sigma \nu} \partial_\lambda g_{\{E\} \{A\}}$.

All other components of the~Riemann tensor that involve exactly two lowercase Greek indices can be obtained from $R^{\{C\}}_{\quad \sigma \{A\} \nu}$ with the~help of symmetries (\ref{Rsymmetries}). For example,
\begin{align}
\begin{split}
\label{riemann_compo4}
& R^{\rho}_{\ \{A\} \sigma \{B\}} = g^{\rho \lambda} R_{\lambda \{A\} \sigma \{B\}} = g^{\rho \lambda} R_{\{A\} \lambda \{B\} \sigma} = g^{\rho \lambda} g_{\{A\} \{C\}} R^{\{C\}}_{\quad \lambda \{B\} \sigma}
\\
& = - g^{\rho \lambda} g_{\{A\} \{C\}} \left( \tfrac{1}{2} g^{\{C\} \{D\}} \widehat{\nabla}_\sigma \partial_\lambda g_{\{D\} \{B\}} + \tfrac{1}{4} \partial_\sigma g^{\{C\} \{D\}} \partial_\lambda g_{\{D\} \{B\}} \right)
\\
& = g^{\rho \lambda} \left( - \tfrac{1}{2} \widehat{\nabla}_\sigma \partial_\lambda g_{\{A\} \{B\}} + \tfrac{1}{4} g^{\{C\} \{D\}} \partial_\sigma g_{\{C\} \{A\}} \partial_\lambda g_{\{D\} \{B\}} \right).
\end{split}
\end{align}

We now have all the~information needed to compute the~``horizontal'' components of the~Ricci tensor $\textbf{ric}_{\mu \nu}$. Applying (\ref{riemann_compo1}) and (\ref{riemann_compo3}) to (\ref{ricci_def}), one has
\begin{align*}
\textbf{ric}_{\mu \nu} & = R^C_{\ \, \mu C \nu} = R^\rho_{\ \mu \rho \nu} + R^{\{C\}}_{\quad \mu \{C\} \nu} = \widetilde{R}^\rho_{\ \mu \rho \nu} + R^{\{C\}}_{\quad \mu \{C\} \nu}
\\
& = \widetilde{\textbf{ric}}_{\mu \nu} - \tfrac{1}{2} g^{\{C\} \{D\}} \widehat{\nabla}_\nu \partial_\mu g_{\{C\} \{D\}} - \tfrac{1}{4} \partial_\nu g^{\{C\} \{D\}} \partial_\mu g_{\{C\} \{D\}}.
\end{align*}

This expression can be put into a~more compact and symmetrical form, namely,
\begin{align}
\label{ricci_compo2}
\textbf{ric}_{\mu \nu} & = \widetilde{\textbf{ric}}_{\mu \nu} - \tfrac{1}{4} \left( \widehat{\nabla}_\mu \widehat{\nabla}_\nu \ln |\breve{g}| + g^{\{A\} \{B\}} \widehat{\nabla}_\mu \widehat{\nabla}_\nu g_{\{A\} \{B\}} \right),
\end{align}
\noindent
where we have introduced $\breve{g} := \det \left( g_{\{A\} \{B\}} \right)$.

Let us now compute the~``mixed'' components of the~Ricci tensor $\textbf{ric}_{\mu \{B\}}$ and $\textbf{ric}_{\{A\} \nu}$. Of course, by the~symmetry of $\textbf{ric}$, it suffices to find the~formula for one of them. One can prove that
\begin{align}
\label{ricci_compo3}
\textbf{ric}_{\mu \{B\}} & = - \tfrac{1}{2} \textbf{c}_{\{B\}\{C\}\{D\}} \partial_\mu g^{\{C\}\{D\}} = \tfrac{1}{2} \partial_\mu \textbf{c}_{\{B\}\{C\}\{D\}} g^{\{C\}\{D\}},
\end{align}
\noindent
where the~second equality is the~direct consequence of (iii).

One proves (\ref{ricci_compo3}) by the~following calculation.
\begin{align*}
\textbf{ric}_{\mu \{B\}} & = R^{C}_{\ \mu C \{B\}} = R^{\{C\}}_{\quad \mu \{C\} \{B\}} + \underbrace{R^{\rho}_{\ \mu \rho \{B\}}}_{= \, 0}
\\
& = \Gamma^S_{\ \mu \{B\}} \Gamma^{\{C\}}_{\quad S \{C\}} - \Gamma^S_{\ \mu \{C\}} \Gamma^{\{C\}}_{\quad S \{B\}} - \textbf{c}^{\quad \quad \ S}_{\{C\}\{B\}} \Gamma^{\{C\}}_{\quad \mu S},
\end{align*}
\noindent
where we have used (\ref{riemann2}) and (\ref{riemann_compo2}). In the~last formula, one can replace, by (\ref{connection_compo2}), the~dummy index $S$ with $\{S\}$ (in each term). Moreover, the~first term now vanishes because, by (\ref{connection_compo2}) and (iv),
\begin{align}
\label{ricci_compo401}
\Gamma^{\{C\}}_{\quad \{S\} \{C\}} = \Gamma^{C}_{\ \{S\} C} = \tfrac{1}{2} g^{CD} \partial_{\{S\}} g_{CD} = 0.
\end{align}

One thus has
\begin{align}
\begin{split}
\label{ricci_compo5}
\textbf{ric}_{\mu \{B\}} & = - \Gamma^{\{S\}}_{\quad \mu \{C\}} \Gamma^{\{C\}}_{\quad \{S\} \{B\}} - \textbf{c}^{\quad \quad \ \, \{S\}}_{\{C\}\{B\}} \Gamma^{\{C\}}_{\quad \mu \{S\}}
\\
& = - \Gamma^{\{S\}}_{\quad \mu \{C\}} \left( \Gamma^{\{C\}}_{\quad \{S\} \{B\}} + \textbf{c}^{\quad \quad \ \, \{C\}}_{\{S\}\{B\}} \right) = - \Gamma^{\{S\}}_{\quad \mu \{C\}} \Gamma^{\{C\}}_{\quad \{B\} \{S\}}
\\
& = -\tfrac{1}{2} g^{\{S\}\{T\}} \partial_\mu g_{\{T\}\{C\}} \Gamma^{\{C\}}_{\quad \{B\} \{S\}} = \tfrac{1}{2} \partial_\mu g^{\{S\}\{T\}} g_{\{T\}\{C\}} \Gamma^{\{C\}}_{\quad \{B\} \{S\}}
\\
& = \tfrac{1}{2} \partial_\mu g^{\{S\}\{T\}} \Gamma_{\{T\} \{B\} \{S\}}
\\
& = \tfrac{1}{4} \partial_\mu g^{\{S\}\{T\}} \left( \textbf{c}_{\{S\}\{B\}\{T\}} + \textbf{c}_{\{T\}\{S\}\{B\}} + \textbf{c}_{\{T\}\{B\}\{S\}} \right),
\end{split}
\end{align}
\noindent
where in one of the~terms we have interchanged the~names of the~indices $\{C\}, \{S\}$ and then used (\ref{asymetria2}), (\ref{connection_compo4}) and (\ref{preconnection_compo6}). To obtain (\ref{ricci_compo3}), it only remains to employ the~symmetry of $g^{\{S\}\{T\}}$ and the~skew-symmetry of $\textbf{c}_{\{T\}\{S\}\{B\}}$ in the~first two indices.

Let us now move to computing the~``inner'' components of the~Ricci tensor $\textbf{ric}_{\{A\} \{B\}}$. One has
\begin{align}
\label{ricci_compo6}
\textbf{ric}_{\{A\} \{B\}} = R^{C}_{\ \{A\} C \{B\}} = R^{\rho}_{\ \{A\} \rho \{B\}} + R^{\{C\}}_{\quad \{A\} \{C\} \{B\}}.
\end{align}

By (\ref{riemann_compo4}), the~first term yields
\begin{align}
\begin{split}
\label{ricci_compo7}
R^{\rho}_{\ \{A\} \rho \{B\}} & = g^{\rho \lambda} \left( - \tfrac{1}{2} \widehat{\nabla}_\rho \partial_\lambda g_{\{A\} \{B\}} + \tfrac{1}{4} g^{\{C\} \{D\}} \partial_\rho g_{\{C\} \{A\}} \partial_\lambda g_{\{D\} \{B\}} \right)
\\
& = - \tfrac{1}{8} g^{\rho \lambda} \left( 3 \widehat{\Delta} g_{\{A\} \{B\}} - g_{\{A\} \{C\}} g_{\{B\} \{D\}} \widehat{\Delta} g^{\{C\} \{D\}} \right),
\end{split}
\end{align}
\noindent
where we have introduced the~Laplace--Beltrami operator $\widehat{\Delta} := g^{\mu \nu} \widehat{\nabla}_\mu \widehat{\nabla}_\nu$ and used the~fact that
\begin{align}
\begin{split}
\label{ricci_compo701}
& g^{\rho \lambda} g^{\{C\} \{D\}} \partial_\rho g_{\{C\} \{A\}} \partial_\lambda g_{\{D\} \{B\}}
\\
& = \tfrac{1}{2} \left( \widehat{\Delta} g_{\{A\} \{B\}} + g_{\{A\} \{C\}} g_{\{B\} \{D\}} \widehat{\Delta} g^{\{C\} \{D\}} \right),
\end{split}
\end{align}
\noindent
which in turn is equivalent to the~following obvious equality:
\begin{align*}
g_{\{A\}\{C\}} \widehat{\Delta} \left( g^{\{C\}\{D\}} g_{\{D\}\{B\}} \right) = 0.
\end{align*}

Computing the~second term on the~right-hand side of (\ref{ricci_compo6}) gives
\begin{align*}
& R^{\{C\}}_{\quad \{A\} \{C\} \{B\}}
\\
& = \Gamma^{S}_{\ \{A\} \{B\}} \Gamma^{\{C\}}_{\quad S \{C\}} - \Gamma^{S}_{\ \{A\} \{C\}} \Gamma^{\{C\}}_{\quad S \{B\}} - \textbf{c}^{\quad \quad \ \, S}_{\{C\}\{B\}} \Gamma^{\{C\}}_{\quad \{A\} S}
\\
& = \Gamma^{\rho}_{\ \{A\} \{B\}} \Gamma^{\{C\}}_{\quad \rho \{C\}} - \Gamma^{\rho}_{\ \{A\} \{C\}} \Gamma^{\{C\}}_{\quad \rho \{B\}} - \Gamma^{\{S\}}_{\quad \{A\} \{C\}} \Gamma^{\{C\}}_{\quad \{S\} \{B\}}
\\
& \quad - \textbf{c}^{\quad \quad \ \, \{S\}}_{\{C\}\{B\}} \Gamma^{\{C\}}_{\quad \{A\} \{S\}}
\\
& = \Gamma^{\rho}_{\ \{A\} \{B\}} \Gamma^{\{C\}}_{\quad \rho \{C\}} - \Gamma^{\rho}_{\ \{A\} \{C\}} \Gamma^{\{C\}}_{\quad \rho \{B\}} - \Gamma^{\{S\}}_{\quad \{A\} \{C\}} \Gamma^{\{C\}}_{\quad \{B\} \{S\}},
\end{align*}
\noindent
where we have used (\ref{riemann2}), (\ref{ricci_compo401}), the~fact that $\textbf{c}^{\quad \quad \ \, \rho}_{\{C\}\{B\}} = 0$ and, in the~last step, we have merged the~last two terms in a~similar fashion as in the~first two lines of (\ref{ricci_compo5}).

Expanding the~expression above using (\ref{connection_compo3}), (\ref{connection_compo4}) and (\ref{connection_compo6}) yields
\begin{align}
\begin{split}
\label{ricci_compo9}
& R^{\{C\}}_{\quad \{A\} \{C\} \{B\}}
\\
& = - \tfrac{1}{4} g^{\rho \lambda} \partial_\lambda g_{\{A\} \{B\}} g^{\{C\} \{D\}} \partial_\rho g_{\{C\} \{D\}} + \tfrac{1}{4} g^{\rho \lambda} \partial_\lambda g_{\{A\} \{C\}} g^{\{C\} \{D\}} \partial_\rho g_{\{B\} \{D\}}
\\
& \quad - \tfrac{1}{4} g^{\{S\}\{T\}} g^{\{C\}\{D\}} \left( \textbf{c}_{\{C\}\{A\}\{T\}} + \textbf{c}_{\{T\}\{C\}\{A\}} + \textbf{c}_{\{T\}\{A\}\{C\}} \right)
\\
& \quad \times \left( \textbf{c}_{\{S\}\{B\}\{D\}} + \textbf{c}_{\{D\}\{S\}\{B\}} + \textbf{c}_{\{D\}\{B\}\{S\}} \right)
\\
& = \tfrac{1}{4} g^{\rho \lambda} g^{\{C\} \{D\}} \left( \partial_\lambda g_{\{A\} \{C\}} \partial_\rho g_{\{B\} \{D\}} - \partial_\lambda g_{\{A\} \{B\}} \partial_\rho g_{\{C\} \{D\}} \right)
\\
& \quad - \tfrac{1}{2} \textbf{c}_{\{A\}}^{ \quad \{C\}\{D\}} \left( \textbf{c}_{\{B\}\{C\}\{D\}} + \textbf{c}_{\{B\}\{D\}\{C\}} \right) + \tfrac{1}{4} \textbf{c}^{\{C\}\{D\}}_{\qquad \ \{A\}} \textbf{c}_{\{C\}\{D\}\{B\}},
\end{split}
\end{align}
\noindent
where in the~last equality we have skipped some tedious, but straightforward, calculations.

Inserting (\ref{ricci_compo7}) and (\ref{ricci_compo9}) into (\ref{ricci_compo6}) and simplifying thus obtained expression with the~help of (\ref{ricci_compo701}) one obtains
\begin{align}
\begin{split}
\label{ricci_compo10}
& \textbf{ric}_{\{A\} \{B\}}
\\
& = - \tfrac{1}{4} \left( \widehat{\Delta} g_{\{A\} \{B\}} - g_{\{A\} \{C\}} g_{\{B\} \{D\}} \widehat{\Delta} g^{\{C\} \{D\}} + g^{\rho \lambda} \partial_\rho \ln | \breve{g} | \partial_\lambda g_{\{A\} \{B\}} \right)
\\
& \quad - \tfrac{1}{2} \textbf{c}_{\{A\}}^{ \quad \{C\}\{D\}} \left( \textbf{c}_{\{B\}\{C\}\{D\}} + \textbf{c}_{\{B\}\{D\}\{C\}} \right) + \tfrac{1}{4} \textbf{c}^{\{C\}\{D\}}_{\qquad \ \{A\}} \textbf{c}_{\{C\}\{D\}\{B\}}.
\end{split}
\end{align}

Finally, let us compute the~curvature scalar $r$. Through applying (\ref{ricci_compo2}) and (\ref{ricci_compo10}) to (\ref{scalar_def}), one obtains
\begin{align*}
r & = g^{AB} \textbf{ric}_{AB} = g^{\mu \nu} \textbf{ric}_{\mu \nu} + g^{\{A\} \{B\}} \textbf{ric}_{\{A\} \{B\}}
\\
& = g^{\mu \nu} \widetilde{\textbf{ric}}_{\mu \nu} - \tfrac{1}{4} g^{\mu \nu} \left( \widehat{\nabla}_\mu \widehat{\nabla}_\nu \ln |\breve{g}| + g^{\{A\} \{B\}} \widehat{\nabla}_\mu \widetilde{\nabla}_\nu g_{\{A\} \{B\}} \right)
\\
& \quad - \, \tfrac{1}{4} g^{\{A\} \{B\}} \left( \widehat{\Delta} g_{\{A\} \{B\}} - g_{\{A\} \{C\}} g_{\{B\} \{D\}} \widehat{\Delta} g^{\{C\} \{D\}} + g^{\rho \lambda} \partial_\rho \ln | \breve{g} | \partial_\lambda g_{\{A\} \{B\}} \right)
\\
& \quad - \, \tfrac{1}{2} \textbf{c}^{\{B\}\{C\}\{D\}} \left( \textbf{c}_{\{B\}\{C\}\{D\}} + \textbf{c}_{\{B\}\{D\}\{C\}} \right) + \tfrac{1}{4} \textbf{c}^{\{C\}\{D\}\{B\}} \textbf{c}_{\{C\}\{D\}\{B\}}
\\
& = \widetilde{r} - \tfrac{1}{4} \left( \widehat{\Delta} \ln |\breve{g}| + g^{\{A\} \{B\}} \widehat{\Delta} g_{\{A\} \{B\}} \right)
\\
& \quad - \, \tfrac{1}{4} \left( g^{\{A\} \{B\}} \widehat{\Delta} g_{\{A\} \{B\}} - g_{\{A\} \{B\}} \widehat{\Delta} g^{\{A\} \{B\}} + g^{\rho \lambda} \partial_\rho \ln |\breve{g}| \partial_\lambda \ln |\breve{g}| \right)
\\
& \quad - \tfrac{1}{4} \textbf{c}^{\{B\}\{C\}\{D\}} \left( \textbf{c}_{\{B\}\{C\}\{D\}} + 2 \textbf{c}_{\{B\}\{D\}\{C\}} \right).
\end{align*}

This formula can be further transformed into various equivalent forms through the~following identities\footnote{Identities (\ref{nice_identity1}--\ref{nice_identity3}) can be easily proven by direct calculations.}:
\begin{align}
\label{nice_identity1}
& g^{\{A\} \{B\}} \widehat{\Delta} g_{\{A\} \{B\}} + g_{\{A\} \{B\}} \widehat{\Delta} g^{\{A\} \{B\}} = - 2 \partial^\mu g^{\{A\} \{B\}} \partial_\mu g_{\{A\} \{B\}},
\\
\label{nice_identity2}
& g^{\{A\} \{B\}} \widehat{\Delta} g_{\{A\} \{B\}} - g_{\{A\} \{B\}} \widehat{\Delta} g^{\{A\} \{B\}} = 2 \widehat{\Delta} \ln |\breve{g}|,
\\
\label{nice_identity3}
& \forall \, f \in C^\infty(M) \quad \ \widehat{\Delta} f + \partial^\mu f \, \partial_\mu f = e^{-f} \widehat{\Delta} e^f,
\end{align}
\noindent
where $\partial^\mu := g^{\mu \nu} \partial_\nu$.

For instance, by employing (\ref{nice_identity2}) and (\ref{nice_identity3}) with $f = \tfrac{1}{2} \ln |\breve{g}|$, one can replace the~content of the~second bracket with $4 \frac{\widehat{\Delta} \sqrt{|\breve{g}|}}{\sqrt{|\breve{g}|}}$. Therefore,
\begin{align}
\begin{split}
\label{ricci_scalar1}
r & = \widetilde{r} - \frac{\widehat{\Delta} \sqrt{|\breve{g}|}}{\sqrt{|\breve{g}|}} - \frac{1}{4} \left( \widehat{\Delta} \ln |\breve{g}| + g^{\{A\} \{B\}} \widehat{\Delta} g_{\{A\} \{B\}} \right)
\\
& \quad - \frac{1}{4}\left( \textbf{c}^{\{B\}\{C\}\{D\}} \left( 2 \textbf{c}_{\{B\}\{D\}\{C\}} + \textbf{c}_{\{B\}\{C\}\{D\}} \right) \right).
\end{split}
\end{align}


For the~sake of convenience, let us introduce the~\emph{structure scalar}
\begin{align*}
\textbf{C} := \textbf{c}^{\{B\}\{C\}\{D\}} \left( 2 \textbf{c}_{\{B\}\{D\}\{C\}} + \textbf{c}_{\{B\}\{C\}\{D\}} \right).
\end{align*}
\noindent
It is in general nontrivial. For instance, in the~case of $G = \mathbb{Z}_2$ it can be expressed as
\begin{align*}
\textbf{C} & = \frac{2}{\breve{g}} \left[ \left(g_{\{1\}\{1\}} \right)^2 + 16 g_{\{12\}\{12\}} g_{\{21\}\{21\}} + 16 g_{\{1\}\{12\}} g_{\{1\}\{21\}} - 8 g_{\{1\}\{1\}} g_{\{12\}\{21\}} \right].
\end{align*}

With the~help of (\ref{ricci_scalar1}), we can write the Einstein--Hilbert action (\ref{action}) more explicitly, namely,
\begin{align*}
S_{EH} = \int \left[ \sqrt{|\breve{g}|} \widetilde{r} - \tfrac{1}{4} \sqrt{|\breve{g}|} \left( \widehat{\Delta} \ln |\breve{g}| + g^{\{A\} \{B\}} \widehat{\Delta} g_{\{A\} \{B\}} + \textbf{C} \right) \right] \sqrt{-\tilde{g}} \, d^N x,
\end{align*}
\noindent
where we have expressed the~determinant of the~metric matrix as the~product of the~determinants of its blocks $|g| = -\tilde{g}|\breve{g}|$. We have also omitted the~divergence term $\widehat{\Delta} \sqrt{|\breve{g}|}$.

One can regard the~theory obtained in this way as an~example of a~scalar-tensor theory\footnote{See e.g. \cite{Clifton} and references therein.} with $n^2(n^2-1)/2$ independent scalar fields arranged into a~(symmetrical and non-degenerate) matrix $(g_{\{A\}\{B\}})$. Let us notice that the~field $\sqrt{|\breve{g}|}$ plays a~special role in this theory.
\\

With the~help of (\ref{ricci_compo2}, \ref{ricci_compo3}, \ref{ricci_compo10}), we can write the generalized Einstein equations $\textbf{ric}_{AB}=0$ for the~model with the~block diagonal metric matrix in the~following form:
\begin{align}
\label{Einstein_eqs1}
& \widetilde{\textbf{ric}}_{\mu \nu} = \tfrac{1}{4} \left( \widehat{\nabla}_\mu \widehat{\nabla}_\nu \ln |\breve{g}| + g^{\{A\} \{B\}} \widehat{\nabla}_\mu \widehat{\nabla}_\nu g_{\{A\} \{B\}} \right),
\\
\label{Einstein_eqs2}
& \textbf{c}_{\{B\}}^{ \quad \{C\}\{D\}} \partial_\mu g_{\{C\}\{D\}} = 0,
\\
\begin{split}
\label{Einstein_eqs3}
& \widehat{\Delta} g_{\{A\} \{B\}} - g_{\{A\} \{C\}} g_{\{B\} \{D\}} \widehat{\Delta} g^{\{C\} \{D\}} + \partial^\mu \ln | \breve{g} | \partial_\mu g_{\{A\} \{B\}}
\\
& \quad = - 2 \textbf{c}_{\{A\}}^{ \quad \{C\}\{D\}} \left( \textbf{c}_{\{B\}\{C\}\{D\}} + \textbf{c}_{\{B\}\{D\}\{C\}} \right) + \textbf{c}^{\{C\}\{D\}}_{\qquad \ \{A\}} \textbf{c}_{\{C\}\{D\}\{B\}}.
\end{split}
\end{align}

Equation (\ref{Einstein_eqs1}) is a~projection of the~generalized Einstein equations onto the~$N$-dimensional space--time $M$. Because it implies that
\begin{align}
\label{Einstein_eqs4}
& \widetilde{r} = \tfrac{1}{4} \left( \widetilde{\Delta} \ln |\breve{g}| + g^{\{A\} \{B\}} \widetilde{\Delta} g_{\{A\} \{B\}} \right),
\end{align}
\noindent
we can equivalently write (\ref{Einstein_eqs1}) in the~form of the~standard Einstein equations with a~certain nonzero energy--momentum tensor
\begin{align}
\begin{split}
\label{Einstein_eqs5}
\widetilde{G}_{\mu \nu} & = \tfrac{1}{4} \left[ \left( \widehat{\nabla}_\mu \widehat{\nabla}_\nu - \tfrac{1}{2} g_{\mu \nu} \widehat{\Delta} \right) \ln |\breve{g}| \right.
\\
& \left. \quad \quad + \, g^{\{A\} \{B\}} \left( \widehat{\nabla}_\mu \widehat{\nabla}_\nu - \tfrac{1}{2} g_{\mu \nu} \widehat{\Delta} \right) g_{\{A\} \{B\}} \right],
\end{split}
\end{align}
\noindent
where $\widetilde{G}_{\mu \nu} := \widetilde{\textbf{ric}}_{\mu \nu} - \tfrac{1}{2} g_{\mu \nu} \widetilde{r}$ is the~``classical'' Einstein tensor.

The appearance of a~nonzero energy--momentum tensor can be regarded as a~realization of the~``matter-out-of-geometry'' mechanism \cite{Heller3} or, more precisely, of the~``scalar-fields-out-of-noncommutative-geometry'' mechanism. Equations (\ref{Einstein_eqs2}) and (\ref{Einstein_eqs3}) should then be regarded as the~equations of state of the~scalar fields under consideration.

The continuity equation satisfied by this energy--momentum tensor can be written in the~form
\begin{align}
\label{continuity_eq}
& \widetilde{\textbf{ric}}_{\mu \nu} \partial^\mu \ln |\breve{g}| + \partial_\nu \widetilde{r} + \tfrac{1}{2} \partial^\mu g^{\{A\} \{B\}} \left( \widehat{\nabla}_\mu \widehat{\nabla}_\nu - g_{\mu \nu} \widehat{\Delta} \right) g_{\{A\} \{B\}} = 0.
\end{align}

To prove (\ref{continuity_eq}), let us act on both sides of (\ref{Einstein_eqs5}) with the~operator $4 \widehat{\nabla}^\mu$. Since $\widetilde{G}$ is a divergence-free tensor, the left-hand side becomes zero and one obtains that
\begin{align*}
& 0 = \widehat{\nabla}^\mu \left[ \left( \widehat{\nabla}_\mu \widehat{\nabla}_\nu - \tfrac{1}{2} g_{\mu \nu} \widehat{\Delta} \right) \ln |\breve{g}| + g^{\{A\} \{B\}} \left( \widehat{\nabla}_\mu \widehat{\nabla}_\nu - \tfrac{1}{2} g_{\mu \nu} \widehat{\Delta} \right) g_{\{A\} \{B\}} \right].
\end{align*}

By expanding the~right-hand side of the~equation above, one readily obtains (\ref{continuity_eq}), because
\begin{align*}
& \widehat{\nabla}^\mu \left[ \left( \widehat{\nabla}_\mu \widehat{\nabla}_\nu - \tfrac{1}{2} g_{\mu \nu} \widehat{\Delta} \right) \ln |\breve{g}| + g^{\{A\} \{B\}} \left( \widehat{\nabla}_\mu \widehat{\nabla}_\nu - \tfrac{1}{2} g_{\mu \nu} \widehat{\Delta} \right) g_{\{A\} \{B\}} \right]
\\
& = \left( \widehat{\Delta} \partial_\nu - \tfrac{1}{2} \partial_\nu \widehat{\Delta} \right) \ln |\breve{g}| + g^{\{A\} \{B\}} \left( \widehat{\Delta} \partial_\nu - \tfrac{1}{2} \partial_\nu \widehat{\Delta} \right) g_{\{A\} \{B\}}
\\
& \quad + \partial^\mu g^{\{A\} \{B\}} \left( \widehat{\nabla}_\mu \partial_\nu - \tfrac{1}{2} g_{\mu \nu} \widehat{\Delta} \right) g_{\{A\} \{B\}}
\\
& = \left[ \widehat{\Delta}, \partial_\nu \right] \ln |\breve{g}| + \tfrac{1}{2} \partial_\nu \widehat{\Delta} \ln |\breve{g}| + g^{\{A\} \{B\}} \left[ \widehat{\Delta}, \partial_\nu \right] g_{\{A\} \{B\}} + \tfrac{1}{2} g^{\{A\} \{B\}} \partial_\nu \widehat{\Delta} g_{\{A\} \{B\}}
\\
& \quad + \tfrac{1}{2} \partial_\nu g^{\{A\} \{B\}} \widehat{\Delta} g_{\{A\} \{B\}} + \partial^\mu g^{\{A\} \{B\}} \left( \widehat{\nabla}_\mu \widehat{\nabla}_\nu - g_{\mu \nu} \widehat{\Delta} \right) g_{\{A\} \{B\}}
\\
& = \widetilde{\textbf{ric}}_{\mu \nu} \partial^\mu \ln |\breve{g}| + g^{\{A\} \{B\}} \widetilde{\textbf{ric}}_{\mu \nu} \partial^\mu g_{\{A\} \{B\}} + \tfrac{1}{2} \partial_\nu \left( \widetilde{\Delta} \ln |\breve{g}| + g^{\{A\} \{B\}} \widetilde{\Delta} g_{\{A\} \{B\}} \right)
\\
& \quad + \partial^\mu g^{\{A\} \{B\}} \left( \widehat{\nabla}_\mu \widehat{\nabla}_\nu - g_{\mu \nu} \widehat{\Delta} \right) g_{\{A\} \{B\}}
\\
& = 2 \widetilde{\textbf{ric}}_{\mu \nu} \partial^\mu \ln |\breve{g}| + 2 \partial_\nu \widetilde{r} + \partial^\mu g^{\{A\} \{B\}} \left( \widehat{\nabla}_\mu \widehat{\nabla}_\nu - g_{\mu \nu} \widehat{\Delta} \right) g_{\{A\} \{B\}},
\end{align*}
\noindent
where we have used (\ref{Einstein_eqs4}) and the~identity\footnote{Identity (\ref{nice_identity4}) can be easily proven in the~Riemann normal coordinates.}
\begin{align}
\label{nice_identity4}
& \forall \, f \in C^\infty(M) \quad \  \left[ \widehat{\Delta}, \partial_\nu \right] f = \widetilde{\textbf{ric}}_{\mu \nu} \partial^\mu f.
\end{align}

Let us present another remarkable equation that can be inferred from Einstein equations (\ref{Einstein_eqs1}--\ref{Einstein_eqs3}). Namely, by (\ref{ricci_scalar1}), (\ref{Einstein_eqs4}) and the~fact that $r=0$ it is true that\footnote{Equation (\ref{Einstein_eqs6}) can also be obtained from (\ref{Einstein_eqs3}) through the~use of the~trace operation with the~help of identities (\ref{nice_identity1}--\ref{nice_identity3}).}
\begin{align}
\label{Einstein_eqs6}
& \widehat{\Delta} \sqrt{|\breve{g}|} + \tfrac{1}{4} \textbf{C} \sqrt{|\breve{g}|} = 0.
\end{align}

Equation (\ref{Einstein_eqs6}), written in this form, resembles the~Klein--Gordon equation. However, the~structure scalar need not be constant, therefore one should interpret it as a~certain potential rather the~the mass term.
\\

Let us finish this section by presenting two explicit Friedmann-like solutions of Einstein equations (\ref{Einstein_eqs1}--\ref{Einstein_eqs3}) in the~simplest case when the~structure group $G$ is equal to $\mathbb{Z}_2$. By a~``Friedmann-like'' solution we mean a~metric, whose ``horizontal'' part $(g_{\mu \nu})$ is of the~Friedmann--Lema\^{\i}tre--Robertson--Walker form and whose remaining components depend only on the~time variable\footnote{We adopt the~system of units in which $c=G=1$.}
\begin{align*}
& g_{\mu \nu} =
\left[\begin{array}{ccccccc}
    -1 & 0 & 0 & 0 \\
    0 & \frac{a^2(t)}{1 - k r^2} & 0 & 0\\
    0 & 0 & a^2(t) r^2 & 0 \\
    0 & 0 & 0 & a^2(t) r^2 \sin^2 \theta
\end{array}\right],
\\
& g_{\{A\} \{B\}} = g_{\{A\} \{B\}}(t),
\end{align*}
\noindent
where $(r,\theta,\phi)$ denote the~reduced-circumference polar coordinates, $a(t)$ is the~scale factor and $k \in \{-1,0,1\}$ is the~curvature constant.

To further simplify calculations, we assume that the~``inner'' part of the~metric matrix $\left( g_{\{A\} \{B\}} \right)$ has the~following form:
\begin{align}
\label{FLRW_metric2}
g_{\{A\} \{B\}}(t) =
\left[\begin{array}{ccc}
    \xi f^2(t) & 0 & 0\\
    0 & 0 & \eta f^2(t)\\
    0 & \eta f^2(t) & 0
  \end{array}\right],
\end{align}
\noindent
where $\xi, \eta$ are nonzero constants and $f$ is a~time-dependent nonvanishing function.

For the~metrics with the~``inner'' part of the~form (\ref{FLRW_metric2}), the~second Einstein equation (\ref{Einstein_eqs2}) is satisfied automatically. The~remaining two equations (\ref{Einstein_eqs1}, \ref{Einstein_eqs3}) amount to the~following overdetermined nonlinear system of ODEs
\begin{align}
\label{FLRW_metric3}
\left\{ \begin{array}{rl}
    \frac{\ddot{a}}{a} + \frac{\ddot{f}}{f} & = 0,
    \\
    a \ddot{a} f + 2 \dot{a}^2 f + 3 a \dot{a} \dot{f} & = - 2 k f,
    \\
    f \ddot{f} a + 2 \dot{f}^2 a + 3 f \dot{f} \dot{a} & = \frac{1}{\eta} a.
  \end{array}\right.
\end{align}
\noindent
together with an~additional algebraical condition that $\xi = 2 \eta$. Note that this system of ODEs is symmetric with respect to the~interchange $(a,k) \leftrightarrow (f,-(2\eta)^{-1})$. Bearing this in mind, we will call both functions $a$ and $f$ ``scale factors''.

It is noteworthy that, by the~third equation of (\ref{FLRW_metric3}), the~function $f$ cannot be constant. Moreover, one can express the~Hubble parameter $H$ in terms of $f$ and its time derivatives as
\begin{align*}
H := \frac{\dot{a}}{a} = \frac{\eta^{-1} - f \ddot{f} - 2 \dot{f}^2}{3 f \dot{f}}.
\end{align*}

The detailed study of system (\ref{FLRW_metric3}) goes far beyond the~scope of this paper and will be addressed in the~future work. Here, we only find the~solutions involving scale factors $a,f$ that are linear
\begin{align*}
a(t) & = a_1 (t - t_0) + a_0,
\\
f(t) & = f_1 (t - t_0) + f_0,
\end{align*}
\noindent
where $a_0,a_1,f_0,f_1,t_0$ are constants. As such, $a$ and $f$ satisfy the~first ODE in (\ref{FLRW_metric3}) trivially.

The remaining two ODEs now imply the~following nonlinear system of \emph{algebraic} equations:
\begin{align*}
\left\{ \begin{array}{rl}
    5 a_1^2 f_1 & = -2k f_1,
    \\
    2 a_1^2 f_0 + 3 a_0 a_1 f_1 & = - 2 k f_0,
    \\
    5 f_1^2 a_1 & = \frac{1}{\eta} a_1,
    \\
    2 f_1^2 a_0 + 3 f_0 f_1 a_1 & = \frac{1}{\eta} a_0.
  \end{array}\right.
\end{align*}

One can show that this system has only two nonzero solutions, one for $k=0$ and another for $k=-1$
\begin{align}
\label{FLRW_sol1}
& \textnormal{for } k=0 && a(t) = a_0, && f(t) = \tfrac{1}{\sqrt{2\eta}} (t - t_0),
\\
\label{FLRW_sol2}
& \textnormal{for } k=-1 && a(t) = \sqrt{\tfrac{2}{5}} (t - t_0), && f(t) = \tfrac{1}{\sqrt{5\eta}} (t - t_0),
\end{align}
\noindent
where $\eta, a_0, t_0$ are constants. Note that $\eta$ cancels out when one substitutes the~above solutions into (\ref{FLRW_metric2}), therefore without any loss of generality one can take $\eta=1$. Note also that for $t=t_0$ the~metric matrix $(g_{AB})$ becomes degenerate.

The first solution (\ref{FLRW_sol1}) describes the~flat Minkowski space--time, although it is \emph{not} a~static solution since $\dot{f} \neq 0$ (and the~previously mentioned degeneracy occurs at $t=t_0$).

The second solution (\ref{FLRW_sol2}) describes a~hyperbolic, linearly expanding universe with the~initial singularity at $t=t_0$. Here the~Hubble parameter $H = (t-t_0)^{-1}$. The~age of the~universe is therefore nothing but the~Hubble time $H_0^{-1}$, where $H_0$ denotes the~Hubble constant\footnote{From the~observational data $H_0^{-1} = (14.42 \pm 0.16) \times 10^9 \textnormal{ yr}$ \cite{Planck}.}.

In the~context of the~Friedmann-like solutions, it might be interesting to see what kind of perfect fluid should be assumed in the~standard Friedmann equations to obtain these particular solutions. Rephrasing the~last statement, we are interested in \emph{what kind of matter--energy is in this case induced by the~(noncommutative) geometry}.

Of course, in the~case of Minkowski solution (\ref{FLRW_sol1}) no matter--energy appears. However, in the~case of hyperbolic solution (\ref{FLRW_sol2}), the~``classical'' Einstein tensor assumes a~nontrivial form
\begin{align*}
& \widetilde{G}_{\mu \nu} = \tfrac{3}{2(t - t_0)^2} \, \textnormal{diag}(3,1,1,1)
\end{align*}

Therefore, the~induced perfect fluid energy density $\rho$ and pressure $p$ are
\begin{align*}
\rho(t) = - \frac{36 \pi}{(t - t_0)^2}, \qquad p(t) = \frac{12 \pi}{(t - t_0)^2}.
\end{align*}

Note that $\rho$ is negative and therefore in this case no classical fluid matches the~induced one. The~equation of state of the~fluid reads here $p = - \tfrac{1}{3} \rho$.

\section{Concluding remarks}
In the~present work, we have studied a~very simplified model of our approach to the~unification of general relativity and quantum mechanics. The~justification of this strategy is that in this simplified model many effects, which in our previous works could only be theoretically indicated, are now explicitly calculated. Although the~model considered in the~present work is strongly simplified, it enjoys some interesting properties that could be illuminating also for more realistic models.
In the~considered model, the~full module of derivations $\textnormal{Der} \, {\cal A}_n$ is locally free of finite rank. This fact has allowed us to construct, by using a~local basis, the~(smooth) geometry of the~gravitational sector in a~computable way. Generalized Einstein equations have been derived with the~help of the~action principle for a~suitably defined analogue of the~Einstein--Hilbert action with no matter terms. When projected onto space--time manifold, the~standard Einstein equations have been recovered plus some extra term, which can be interpreted as an~energy--momentum tensor arising from noncommutativity, and the~components of this tensor as a~set of scalar fields together with their equations of state. We have been able to illustrate this ``matter-out-of-geometry'' mechanism by studying two explicit solutions to the~generalized Einstein equations.

In the~present work, we have considered the~gravitational sector of our model in the~case of a~finite structure group. The~quantum sector of this model was considered in \cite{HPS05,finite1}. Let us take a~quick look at its main features. To construct the~quantum sector one considers the~regular representation of the~algebra ${\cal A}_n$
\begin{align*}
\pi_p: {\cal A}_n \rightarrow {\cal B}({\cal H}_p)
\end{align*}
\noindent
in the~collection of Hilbert spaces ${\cal H}_p := L^2(\Gamma^p)$.
This representation establishes an~isomorphism between ${\cal A}_n$ and ${\cal M}_0 := \pi({\cal A}_n)$ where, for any $a \in {\cal A}_n$, $\pi(a) := \left( \pi_p(a) \right)_{p \in M \times G}$.
This isomorphism allows us to transfer geometry of the~gravitational sector to the~quantum sector. It is remarkable that the~geometry, as transferred to ${\cal M}_0$, has a~strong probabilistic flavor. Indeed, the~operators $\pi(a)$ are random operators \cite{Connes}.
To obtain the~``full'' quantum sector, we complete ${\cal M}_0$ to a~von Neumann algebra ${\cal M} := {\cal M}_0^{\prime \prime}$, where the~operators $(\pi_p(a))_{p \in M \times G}$ act in the~Hilbert space given by the~direct integral \cite{HPS05}
\begin{align*}
{\cal H} := \int\limits_{\bigoplus\limits_{p \in M \times G}} {\cal H}_p \, d\mu(p).
\end{align*}
The~isomorphism ${\cal A}_n \cong {\cal M}_0$ does not extend to ${\cal M}$. This shows that geometric methods have a~limited range in the~quantum sector. This sector will be studied, in a~more detailed way, in a~forthcoming work.

However, one should remember that the~model with a~finite structure group is to be regarded as a~step towards constructing more physically realistic models.

\end{document}